\documentclass[aps,pra,twocolumn,superscriptaddress, amsmath,amssymb,10pt,longbibliography]{revtex4-2}
\usepackage[colorlinks=true,linkcolor=blue,urlcolor=blue,citecolor=blue,pdfusetitle]{hyperref}
\usepackage{graphicx}
\usepackage{tikz}
\usepackage{xcolor}
\usepackage{soul}
\usepackage{orcidlink}
\usepackage{float}
\usepackage{libertine} 
\usepackage{siunitx}
\usetikzlibrary{shapes.arrows, decorations.pathmorphing, arrows.meta}

\usepackage{braket}
\usepackage{physics}

\begin{document}

\nocite{apsrev42Control}

\title{Quantum Monte Carlo study of systems interacting via long-range interactions mediated by a cavity}

\author{Marta Dom\'{i}nguez-Navarro~\orcidlink{0009-0003-3931-0321}}
\email{marta.dominguez.navarro@upc.edu}
\affiliation{Departament de F{\'i}sica Qu{\`a}ntica i Astrof{\'i}sica, Facultat de F{\'i}sica, Universitat de Barcelona, E-08028 Barcelona, Spain.}
\affiliation{Institut de Ci{\`e}ncies del Cosmos, Universitat de Barcelona, ICCUB, Mart{\'i} i Franqu{\`e}s 1, E-08028 Barcelona, Spain.}

\author{Abel Rojo-Franc\`{a}s~\orcidlink{0000-0002-0567-7139}}
\email{abel.rojo@oist.jp}
\affiliation{Departament de F{\'i}sica Qu{\`a}ntica i Astrof{\'i}sica, Facultat de F{\'i}sica, Universitat de Barcelona, E-08028 Barcelona, Spain.}
\affiliation{Institut de Ci{\`e}ncies del Cosmos, Universitat de Barcelona, ICCUB, Mart{\'i} i Franqu{\`e}s 1, E-08028 Barcelona, Spain.}
\affiliation{Quantum Systems Unit, Okinawa Institute of Science and Technology Graduate University,\\Onna, Okinawa 904-0495, Japan.}

\author{Bruno Juli\'{a}-D\'{i}az~\orcidlink{0000-0002-0145-6734}}
\email{bruno@fqa.ub.edu}
\affiliation{Departament de F{\'i}sica Qu{\`a}ntica i Astrof{\'i}sica, Facultat de F{\'i}sica, Universitat de Barcelona, E-08028 Barcelona, Spain.}
\affiliation{Institut de Ci{\`e}ncies del Cosmos, Universitat de Barcelona, ICCUB, Mart{\'i} i Franqu{\`e}s 1, E-08028 Barcelona, Spain.}

\author{Grigori E. Astrakharchik~\orcidlink{0000-0003-0394-8094}}
\email{grigori.astrakharchik@upc.edu}
\affiliation{Departament de F\'isica, Universitat Polit\`ecnica de Catalunya, 
Campus Nord B4-B5, E-08034 Barcelona, Spain.}

\begin{abstract}
We study one-dimensional quantum gases in continuous space with cavity-mediated infinite-range interactions using the variational and the diffusion Monte Carlo methods. Starting from the exact two-body solution, we construct a non-translationally invariant Jastrow wave function that accurately captures the spatial structure induced by the cavity field and provides an efficient many-body ansatz for both bosonic and fermionic systems.
We analyze properties of three characteristic quantum systems, subject to long-range interactions: (i) ideal Bose gas, (ii) interacting Bose gas, (iii) ideal Fermi gas.
In the absence of short-range interactions, we identify a crossover from a stable, weakly modulated phase realized at finite particle number for repulsive interactions to a self-organized state for attractive interactions, marked by clustering, loss of superfluidity, and, at fixed interaction strength, the absence of a thermodynamic limit.
Introducing short-range repulsion, either through contact interactions or fermionic statistics, leads to the formation of a mesoscopic gas-like regime that disappears in the thermodynamic limit.
A qualitative phase diagram is proposed to illustrate the combined effects of short- and long-range interactions, highlighting the emergence of distinct regimes with characteristic structural properties. Applying the Kac prescription, we obtain a well-defined thermodynamic limit and show that the periodic modulation found for a finite number of particles in the repulsive regime is a finite-size artifact, while the self-organized state survives with its onset shifted to substantially stronger coupling.
\end{abstract}

\maketitle

\section{Introduction}

The rapid development of quantum simulators has transformed the study of complex many-body phenomena by enabling their realization in highly controllable experimental platforms. Ultracold atomic gases~\cite{greinerQuantumphasetransition2002} and trapped ions~\cite{porras_effective_2004} offer unprecedented control over system parameters, facilitating the engineering of tailored quantum Hamiltonians and the exploration of exotic quantum phases~\cite{lewensteinUltracoldatomicgases2007, blochQuantumsimulationswith2017}.

While short-range interactions are highly tunable and have been extensively studied, both theoretically~\cite{Bloch2008, Chin2010} and experimentally~\cite{jakschColdBosonicAtoms1998}, 
more recently, long-range interactions have attracted sustained interest  
due to their distinct behavior and their capability of creating 
different types of orders simultaneously~\cite{boninsegniColloquimSupersolidsWhat2012}. 
Theoretical studies have shown that long-range interactions can lead to phenomena such as supersolidity and non-equilibrium phase transitions, motivating ongoing efforts to explore their role in quantum many-body systems~\cite{lahayeThePhysicsOf2009, baranovCondensedMatterTheory2012}.
Among the various schemes proposed to realize long-range interactions in quantum gases, high-finesse optical cavities offer a potent and versatile platform~\cite{Ritch13, Mivehvar2021}. In these setups, atoms coherently scatter photons into a common cavity mode, resulting in effective infinite-range interactions mediated by the light field. These interactions are highly tunable in both strength and symmetry and have been shown to drive phenomena such as self-organization~\cite{nagy_self-organization_2008}, collective friction~\cite{Vukics_2005, black_observation_2003}, and symmetry-breaking phase transitions~\cite{gopalakrishnanAtomLightCristallization2010}.

The first experimental realization of cavity-induced self-organization was achieved by Baumann \textit{et al.}~\cite{baumann_dicke_2010}, who observed the Dicke quantum phase transition in a Bose-Einstein condensate coupled to an optical cavity. Shortly after, Mottl \textit{et al.}~\cite{mottlRotontypeModeSoftening2012} observed roton-like mode softening in the excitation spectrum of a similar system. The divergence of the susceptibility at finite momentum was interpreted as a precursor to crystalline order, and such mode softening has been proposed as a possible route to supersolidity, where superfluid order coexists with long-range diagonal order (i.e., density modulation). The scope of cavity QED has since broadened considerably, including the realization of multimode cavities~\cite{landig_quantum_2016}, dynamically tunable optical lattices~\cite{klinder_nonequilibrium_2015}, and programmable optomechanical arrays of levitated particles~\cite{vijayan_cavity-mediated_2024}.

The extension to fermionic systems has also attracted substantial theoretical attention, with early work predicting a superradiant phase transition analogous to the bosonic case, modified by Pauli blocking and the geometry of the Fermi surface~\cite{Keeling2014, Piazza2014, ChenYu2014}. This phenomenology has since been observed experimentally in a quantum-degenerate Fermi gas~\cite{Zhang2021} and, independently, in a strongly interacting unitary Fermi gas exhibiting genuine density-wave order~\cite{helsonDensitywaveOrderingUnitary2023}.

In addition to experimental progress, theoretical studies aim to provide a comprehensive understanding of cavity-mediated long-range interactions. Effective models describing these systems often involve nonlocal and nonlinear couplings~\cite{gopalakrishnanAtomLightCristallization2010, nagy_self-organization_2008, dogra_phase_2016}, which might require non-standard analytical and numerical approaches. In particular, mean-field and perturbative methods are generally inadequate in strongly interacting regimes. In this context, numerical techniques, especially Quantum Monte Carlo (QMC) methods, offer a powerful non-perturbative framework for investigating ground state properties and correlation functions in correlated quantum gases. Previous works have employed both computationally exact methods, such as Matrix Product States~\cite{Halati2020}, and Monte Carlo methods~\cite{dogra_phase_2016, karpov_light-induced_2022} to study cavity-mediated long-range interactions in lattice systems, specifically within extended Bose-Hubbard models. Continuous-geometry results have been obtained with MCTDH-X in Ref.~\cite{Molignini_2022}; to the best of our knowledge, however, no QMC study in continuous geometry has been reported to date.

The main goal of this work is to study the ground state properties of a one-dimensional quantum gas with cavity-mediated long-range interactions, based on the ideas proposed in Ref.~\cite{mottlRotontypeModeSoftening2012}, by using a similar form of periodic infinite-range interactions. To obtain the ground state properties of the many-body problem, we use QMC techniques, that is, variational and diffusion Monte Carlo methods. We obtain accurate estimates of experimentally relevant observables such as the ground state energy, density profiles, and static structure factor. We perform an analysis of how the cavity-mediated interactions affect the following three fundamental classes of quantum systems: (i) the ideal Bose gas, (ii) a weakly interacting Bose gas, (iii) a spinless ideal Fermi gas. This comparative approach enables us to identify how long-range correlations manifest differently depending on the underlying quantum statistics.

The structure of this work is as follows: In Sec.~\ref{sec:physical_model}, we introduce the physical context of the cavity setup and present the effective model Hamiltonian, together with the approximations used to reduce the system to a one-dimensional description, as well as outline the numerical techniques used, with particular attention to the construction of the trial wave functions and the implementation of the Monte Carlo algorithms. Building on these foundations, we present our first results in Sec.~\ref{sec:two_particle}, where we analyze the simple case of two particles to get a first look into the effects of the cavity-mediated interaction in a quantum system and also to use for the construction of the trial wave function in the many-body case. In Sec.~\ref{sec:many_body_bosons}, we extend the analysis to many-body systems of non-interacting bosons, characterizing the ground state energy and density profiles for different numbers of particles. To further study the stability and properties of the long-range interaction in other systems, in Sec.~\ref{sec:short_range_and_fermi}, we incorporate additional short-range physics, contact interactions for bosons, and Pauli's exclusion constraint for fermions, and study how these contributions affect the system’s behavior in the presence of the long-range cavity potential. Finally, in Sec.~\ref{sec:kac-scaling} we address the proper thermodynamic limit of the system using the Kac prescription~\cite{Kac1963}, rescaling the long-range interaction with the total particle number to obtain well-defined, $N$-independent values of the energy per particle and the density profile as $N\to\infty$. Section~\ref{sec:conclusion} then summarizes the main results obtained and future directions.

\section{System and Methods}
\label{sec:physical_model}
In this section, we present the physical model that forms the basis of our numerical study. We begin by reviewing the key features of the experimental setup of Ref.~\cite{mottlRotontypeModeSoftening2012}, where ultracold atoms are coupled to a single-mode optical cavity to realize tunable long-range interactions, as illustrated schematically in Fig.~\ref{fig:setup}. We then introduce the full microscopic Hamiltonian describing the light-matter system, followed by a dimensional reduction leading to a simplified one-dimensional model in dimensionless units. This reduced model serves as the starting point for all subsequent numerical simulations.

Furthermore, we outline the numerical methods used throughout this work. Our approach consists of three numerical methods: imaginary time evolution (ITE) to extract the two-body wave function; the variational Monte Carlo (VMC) method to explore the many-body system using a variational ansatz based on the two-body solution; and the diffusion Monte Carlo (DMC) method to calculate exact estimates, within statistical error, of the energy in several cases and to validate our variational results. The full implementation of all numerical methods used is available in a public GitHub repository~\cite{MartaDominguez2025}.

\subsection{System Hamiltonian}
\label{sec:system_hamiltonian}

\begin{figure}[t]
\centering
\begin{tikzpicture}[scale=0.85]

\fill[gray!20, opacity=0.5]
    plot[domain=-2.8:2.8, samples=40, smooth]
        (\x, { 0.18*sqrt(1 + (\x/0.6)^2) + 0.5}) --
    plot[domain=2.8:-2.8, samples=40, smooth]
        (\x, {-0.18*sqrt(1 + (\x/0.6)^2) - 0.5}) -- cycle;
\draw[gray!60, thick, domain=-2.8:2.8, samples=40, smooth]
    plot (\x, { 0.18*sqrt(1 + (\x/0.6)^2) + 0.5});
\draw[gray!60, thick, domain=-2.8:2.8, samples=40, smooth]
    plot (\x, {-0.18*sqrt(1 + (\x/0.6)^2) - 0.5});

\fill[teal!30!cyan, opacity=0.1] (0,0) ellipse (2.4 and 0.8);
\draw[teal!60!cyan, thick]   (0,0) ellipse (2.4 and 0.8);

\draw[gray!80!black, very thick, domain=-2.8:2.8, samples=80, smooth]
    plot (\x, {0.4 * cos(deg(9.3 * \x))});

\foreach \x in {-1.0, -0.35, 0.35, 1.0}{
    \fill[pink!40!red, opacity=0.4] (\x, 0) circle (0.20);
    \fill[pink!10!red]              (\x, 0) circle (0.09);
}

\fill[gray!40]
    (-3.2,-1.55) -- (-2.5,-1.55)
    .. controls (-2.75,-0.7) and (-2.75,0.7) ..
    (-2.5, 1.55) -- (-3.2, 1.55) -- cycle;
\draw[black!70, thick]
    (-3.2,-1.55) -- (-2.5,-1.55)
    .. controls (-2.75,-0.7) and (-2.75,0.7) ..
    (-2.5, 1.55) -- (-3.2, 1.55) -- cycle;

\fill[gray!40]
    ( 3.2,-1.55) -- ( 2.5,-1.55)
    .. controls ( 2.75,-0.7) and ( 2.75,0.7) ..
    ( 2.5, 1.55) -- ( 3.2, 1.55) -- cycle;
\draw[black!70, thick]
    ( 3.2,-1.55) -- ( 2.5,-1.55)
    .. controls ( 2.75,-0.7) and ( 2.75,0.7) ..
    ( 2.5, 1.55) -- ( 3.2, 1.55) -- cycle;

\draw[-{Triangle[width=45pt, length=9pt]},
      teal!60!gray,
      line width=38pt]
      (0, -1.8) -- (0, -1.2);
\node[below, font=\small] at (0, -1.85) {Pump};
\node[right, font=\small] at (1., -1.6) {$\eta,\,\omega_p$};

\draw[-{Triangle[width=45pt, length=9pt]},
      teal!60!gray,
      line width=38pt]
      (0, 1.8) -- (0, 1.2);

\node[font=\small, teal!80!cyan] at (2.1, 0.7) {BEC};

\draw[
    -{Stealth[length=7pt, width=5pt]},
    black, thick, decorate,
    decoration={snake, amplitude=2.5pt, segment length=6pt, post length=8pt}
] (3.25, 0) -- (4.2, 0);
\node[right, font=\small, black] at (4.2, 0) {$\kappa$};

\begin{scope}[shift={(-2.8, -1.9)}, scale=0.6]
    \draw[thick, black] (0,0) circle (0.15);
    \fill[black]        (0,0) circle (0.05);
    \node[above left, font=\footnotesize, black] at (-0.1, -0.1) {$y$};
    \draw[->, thick, black] (0,0) -- (0.75, 0);
    \node[right, font=\footnotesize, black] at (0.78, 0) {$x$};
    \draw[->, thick, black] (0,0) -- (0, -0.75);
    \node[below, font=\footnotesize, black] at (0, -0.78) {$z$};
\end{scope}

\end{tikzpicture}
\caption{Schematic representation of the typical cavity-QED setup, similar to that of Ref.~\cite{mottlRotontypeModeSoftening2012}. A BEC is placed inside a high-finesse optical cavity and illuminated by a transverse pump laser of frequency $\omega_p$ and two-photon Rabi frequency $\eta$. Coherent atom-photon scattering populates the cavity standing-wave mode. Photons leak out at rate $\kappa$. The cavity axis ($x$ direction) defines the geometry of the effective 1D model.}
\label{fig:setup}
\end{figure}
We consider a system inspired by the experimental setup of Ref.~\cite{mottlRotontypeModeSoftening2012}, where a Bose-Einstein condensate (BEC) is coupled to a single-mode optical cavity. The theoretical framework follows Refs.~\cite{maschlerColdAtomDynamics2005, maschlerUltracoldAtomsOptical2008}, which generalize the Jaynes-Cummings model in a rotating-wave and dipole approximation to many-body systems of ultracold atoms by treating both the cavity field and the atomic motion quantum mechanically. The model assumes that it is in the limit of a strongly damped cavity, where the cavity field dynamics can be adiabatically eliminated, making its dynamics fully follow only the atomic configuration. 
Beyond this approximation, deviations have been reported even in one-dimensional systems~\cite{Halati2020}.

Under adiabatic elimination of the electronically excited states and assuming the system reaches a steady state~\cite{maschlerColdAtomDynamics2005, maschlerUltracoldAtomsOptical2008, mottlRotontypeModeSoftening2012}, 
the system can be described by an effective Hamiltonian of the form
\begin{align}
\label{eq:eff_hamiltonian}
\hat{H}_{\text{eff}} = \hat{H}_A + \hat{V}_{\text{int}},
\end{align}
where $\hat{H}_A$ accounts for the atomic kinetic energy, external trapping, and contact interactions, and $\hat{V}_{\text{int}}$ describes the cavity-mediated long-range interactions.

The interaction potential takes the form
\begin{align}
\label{eq:long_range_int}
\hat V_{\mathrm{int}}
= V_0 &\int d^3r\, d^3r'\;
\hat\Psi^\dagger(\mathbf r)\hat\Psi^\dagger(\mathbf r') \cos(k_\text{L} x)\cos(k_\text{L} z)\nonumber\\
&\times \cos(k_\text{L} x')\cos(k_\text{L} z')\hat\Psi(\mathbf r)\hat\Psi(\mathbf r'),
\end{align}
where $\hat{\Psi}(\mathbf{r})$ is the bosonic field operator, and $k_\text{L} = 2\pi / \lambda$ is the wavevector associated with the pump laser. The interaction strength reads
\begin{align}
    \label{eq:V0_definition}
    V_0 \approx \hbar \frac{\eta^2}{\Delta_c - U_0 \mathcal{B}_0},
\end{align}
with a quadratic dependence on the two-photon Rabi frequency $\eta$, which characterizes the coherent scattering rate into the cavity mode and depends on pump laser intensity. The detuning $\Delta_c$ is the frequency difference between the pump laser and cavity resonance. The single-photon light shift is $U_0 = g_0^2 /\Delta_a$, where $g_0$ is the vacuum Rabi frequency and $\Delta_a$ is the atomic detuning. The bunching parameter is defined explicitly as
\begin{align}
    \label{eq:bunching_parameter}
    \mathcal{B}_0 \equiv \langle \hat{\mathcal{B}} \rangle = \int d^3r \, \langle \hat{\Psi}^\dagger(\mathbf{r}) \cos^2(k_\text{L} x) \hat{\Psi}(\mathbf{r}) \rangle,
\end{align}
which quantifies the spatial overlap between the atomic density and the pump intensity pattern.

Besides the interaction potential, the effective Hamiltonian contains an atomic contribution given by
\begin{align}
\label{eq:atomic_hamiltonian_text}
\hat{H}_A = \int d^3 \mathbf{r} \, \hat{\Psi}^\dagger(\mathbf{r}) \biggl[ -\frac{\hbar^2 \nabla^2}{2m} &+ V_p \cos^2(k_\text{L} z) \nonumber \\
&+ \frac{g}{2} \hat{\Psi}^\dagger(\mathbf{r}) \hat{\Psi}(\mathbf{r}) \biggl] \hat{\Psi}(\mathbf{r}),
\end{align}
where $V_p$ is the amplitude of the static optical lattice along $z$, and $g = 4\pi \hbar^2 a / m$ is the coupling constant, characterizing the strength of interactions, described by the $s$-wave scattering length $a$. Experimentally, both $V_p$ and $V_0$ depend on the two-photon Rabi frequency $\eta$ and are therefore not independent. In general, increasing $V_p$ also implies increasing $V_0$, unless compensated through adjustment of the cavity detuning $\Delta_c$ and atomic detuning $\Delta_a$. Nevertheless, in the present theoretical study, $V_0$ is treated as a free parameter of the effective one-dimensional model, allowing systematic exploration of different interaction regimes.

The total effective Hamiltonian thus retains the kinetic and contact interaction terms of a conventional atomic system but also incorporates a nonlocal interaction that is both periodic and long-ranged. Crucially, this interaction depends explicitly on the absolute position of atoms rather than their relative separations, thus breaking translational invariance. Due to the spatial structure inherited from the cavity and pump modes, atoms separated by integer multiples of the cavity wavelength interact identically, highlighting the infinite-range and periodic nature of the induced interactions. 

\subsection{One-dimensional Geometry} 
\label{sec:1d_hamiltonian}

The full three-dimensional Hamiltonian Eq.~(\ref{eq:eff_hamiltonian}), derived above, serves as the starting point for our analysis. To make the simulations of the system properties more numerically tractable, we consider a simplified version of the many-body problem, in which the atomic motion is restricted to one dimension ($x$-axis). This approximation corresponds to the limit of a deep transverse optical lattice. In particular, a large lattice height $V_p$ in Eq.~\eqref{eq:atomic_hamiltonian_text} freezes the atomic motion along the $z$ direction, so that the separation between transverse energy levels becomes large and they cannot be excited, effectively reducing the system to two dimensions. A strictly one-dimensional geometry is then obtained by assuming an additional strong confinement in the remaining transverse direction, such that atomic motion is restricted to the $x$ axis only.

The resulting effective one-dimensional Hamiltonian reads
\begin{align}
\label{eq:1D_hamiltonian_dimensionless}
H_{\text{1D}} = \sum_{i=1}^{N} \left( -\frac{\hbar^2}{2m} \frac{\partial^2}{\partial x_i^2} \right) &+ \sum_{i<j} V_0 \cos\left(k_{\text{L}} x_i\right) \cos\left(k_{\text{L}} x_j\right) \nonumber\\
&+ \sum_{i<j} g \, \delta(x_i - x_j),
\end{align}
where $N$ is the number of atoms, $g = -2\hbar^2 / (m a_{\text{1D}})$ is the effective one-dimensional coupling constant, and $a_{\text{1D}}$ is the one-dimensional $s$-wave scattering length. The second term in Eq.~(\ref{eq:1D_hamiltonian_dimensionless}) describes a cavity-mediated long-range interaction with amplitude $V_0$ and the wavevector $k_\text{L}$. To model repulsive contact interactions, we choose a positive coupling constant $g > 0$ which corresponds to a negative scattering length, $a_{\text{1D}} < 0$.
The cavity-mediated interaction is of infinite range and, unlike the contact interaction, breaks translational invariance, imposing a spatial periodicity given by $k_\text{L}$.

The underlying optical lattice sets natural units for the energy and length to compare with.
We choose the cavity wavelength $\lambda$ as the unit of length and define the natural energy scale as
\begin{align*}
\varepsilon = \frac{\hbar^2}{m \lambda^2},
\end{align*}
which is proportional to the recoil energy.

\subsection{Exact wave function for two particles}
\label{sec:two_body_wf_ITE}

To obtain the ground state energy $E_0$ and wave function $\psi_2(x_1, x_2)$, we solve the Schr\"{o}dinger equation using the ITE method for two particles interacting via the cavity-mediated potential given by the second term of Eq.~\eqref{eq:1D_hamiltonian_dimensionless} in a one-dimensional box with periodic boundary conditions. 
We evolve the system in imaginary time by replacing $t \rightarrow -i\tau/\hbar$ in the time-dependent Schr\"{o}dinger equation. 
We start from a continuous normalized trial wave function $\psi(x_1, x_2,\tau=0)$ and numerically solve the Schr\"{o}dinger differential equation.

As the imaginary time increases, higher-energy components decay faster than the ground state component, and the wave function is projected to the ground state of the Hamiltonian (up to normalization):
\begin{align}
\psi(x_1, x_2, \tau) \xrightarrow{\tau \to \infty} \psi_2(x_1, x_2) e^{-E_0 \tau}
\end{align}

The resulting wave function is sampled on a discrete spatial grid and is interpolated using cubic B-splines. Obtaining the exact solution of the two-body problem plays two major roles in our analysis: it provides physical insight into the two-body correlations induced by the cavity interaction, and it serves as a building block for constructing the many-body trial wave function we used in variational and diffusion Monte Carlo simulations.

\subsection{Trial wave function for many-body systems}
\label{sec:trial_wf}
To construct a trial wave function suitable for studying the many-body system, the standard approach is to use the pair-product Jastrow ansatz~\cite{McMillan}, which encodes pairwise correlations in homogeneous quantum systems,
\begin{align}
\Psi_{\text{Jastrow}}(\mathbf{x}) = \prod_{i<j} f(|x_i-x_j|),
\label{Eq:Jastrow:homogeneous}
\end{align}
where the pairwise function $f(|x_i-x_j|)$ depends only on the interparticle distance $|x_i-x_j|$, assuming translational invariance of the interaction potential. However, in our case, the cavity-mediated interaction explicitly breaks translational invariance. Thus the standard Jastrow form~(\ref{Eq:Jastrow:homogeneous}) fails to capture the physics of the interaction adequately. Instead, we adopt a position-resolved ansatz based on the two-body wave function $\psi_2(x_i, x_j)$ that depends on the positions of both particles
\begin{align}
    \label{eq:position_ansatz}
    \Psi_{\text{position}}(\mathbf{x})=\prod_{i<j} f(x_i,x_j).
\end{align}
This allows us to accurately capture the position-dependent correlations induced by the periodic long-range interaction.

\subsubsection{Bosons with long-range interaction}

 We use the wave function $\psi_2(x_i, x_j)$ of the exact (numerical) solution of the Schr\"{o}dinger problem to build a trial many-body wave function for long-range interactions. Following the spirit of Jastrow-type wave functions, we extend the pairwise structure by taking the product over all pairs of particles,
\begin{align}
\label{eq:trial_wavefunction_initial}
\Psi^{\text{(B)}}_{\text{LR}}(x_1, \ldots, x_N) = \prod_{i<j} \psi_2(x_i, x_j).
\end{align}
This form allows us to retain the spatial correlations induced by the cavity-mediated interaction. Since the two-particle wave function depends on the positions of both particles rather than just their relative distance, Eq.~\eqref{eq:trial_wavefunction_initial} can capture the non-translationally-invariant structure for the many-body case.

However, wave function~\eqref{eq:trial_wavefunction_initial} only includes the cavity interaction. To be able to explore more relevant physical regimes, we consider two modifications of this wave function to account for (i) short-range contact interactions, (ii) fermionic statistics. These extra factors are multiplied into the wave function and can be added or not depending on the parameters we choose.

\subsubsection{Bosons with short- and long-range interaction}
To incorporate short-range repulsion, which is represented in Eq.~\eqref{eq:1D_hamiltonian_dimensionless} by a delta function, we propose the following expression for the trial wave function
\begin{align}
\label{eq:contact_term}
\Psi^{\text{(B)}}_{\text{SR+LR}}(x_1, \dotsc , x_N) = &\prod_{i<j} \psi_2(x_i, x_j) \nonumber \\ &\times \prod_{i<j} f_{2B}(|x_i-x_j|),
\end{align}
where $f_{2B}(|x|) = \cos\left[ k_c \left(|x| - \frac{L}{2}\right) \right]$, which is chosen to satisfy
the Bethe-Peierls boundary condition 
\begin{equation}
    \frac{f_{2B}'(x)}{f_{2B}(x)}\bigg|_{x\rightarrow 0} =-\frac{1}{a_{1D}},
\end{equation}
which allows us to include the effect of contact interactions without having to deal with the singular nature of the delta potential directly \cite{BethePeierls35}. As well, $f_{2B}(x)$ is chosen to satisfy the periodic boundary conditions, as $f'_{2B}\left(\frac{L}{2}\right) = 0$. $L$ corresponds to the size of the 1D box and $k_c$ is determined by the one-dimensional scattering length $a_\text{1D}$ through the Bethe-Peierls boundary condition.

\subsubsection{Fermions with long-range interaction}

In addition to short-range repulsion, we also explore the effect of quantum statistics. More specifically, we compare our bosonic system to its fermionic counterpart. 
According to Bose-Einstein statistics, the many-body wave function must be symmetric under the exchange of any two particles. Indeed, wave functions~\eqref{eq:trial_wavefunction_initial} and~\eqref{eq:contact_term} preserve bosonic symmetry, that is
\begin{align}
    \Psi^{\text{(B)}}_{\text{LR}}(\dotsc,x_i, \dotsc,x_j,\dotsc) &=  \Psi^{\text{(B)}}_{\text{LR}}(\dotsc,x_j, \dotsc,x_i,\dotsc),  \nonumber\\
    \Psi^{\text{(B)}}_{\text{SR+LR}}( \dotsc,x_i, \dotsc,x_j,\dotsc) &=  \Psi^{\text{(B)}}_{\text{SR+LR}}(\dotsc,x_j, \dotsc,x_i,\dotsc),
\end{align}
remain the same under the exchange of any two particles $i$ and $j$, where $N$ is the total number of particles in the system.

However, for Fermi-Dirac statistics, the many-body wave function must be antisymmetric under two-particle exchange, which implies that
\begin{align}
\Psi^{\text{(F)}}_{\text{LR}} (\dotsc,x_i, \dotsc,x_j,\dotsc) = - \Psi^{\text{(F)}}_{\text{LR}} (\dotsc,x_j, \dotsc,x_i,\dotsc).
\end{align}
To introduce the antisymmetrization of the wave function, we consider the ground state wave function of an ideal Fermi gas in the absence of interaction as a Slater determinant, which in one dimension takes the form of a Vandermonde determinant and can be written as a pair product
\begin{align}
\Psi^{\text{(F)}}(x_1, \dotsc, x_N) &=  \frac{1}{\sqrt{N!}} \det \left[ e^{i k_j x_i} \right]_{i,j=1}^N \nonumber \\ &= \text{const} \times \prod_{i<j} \sin\left[\frac{\pi}{L}(x_i - x_j) \right],
\end{align}
where $k_j = \frac{2\pi}{L} n_j$, and the $n_j$ are integers labeling the $N$ lowest-energy single-particle states.

To take the long-range interactions into account, we use the following wave function to describe fermions, 
\begin{align}
\label{eq:fermi-stats_term}
\Psi_{\text{LR}}^{\text{(F)}}(x_1, \dotsc ,x_N) \! = \! \prod_{i<j} \! \psi_2  (\!x_i, x_j\!) \! \prod_{i<j}  \sin \! \left[ \!\frac{\pi}{L}(\!x_i \! - \! x_j\!) \! \right] \! .
\end{align}
This wave function vanishes whenever any two particles are at the same position, enforcing the nodal structure required for a system of spinless fermions.

The product structure of Eqs.~\eqref{eq:contact_term} and~\eqref{eq:fermi-stats_term} incorporates the relevant physical ingredients. 
The factor
$\prod_{i<j} \psi_2(x_i, x_j)$ 
captures the correlations induced by the cavity-mediated potential. In Eq.~\eqref{eq:contact_term}, the additional factor
$\prod_{i<j} f_{2B}(|x_i-x_j|)$
encodes short-range two-body correlations via the Bethe-Peierls boundary condition. In Eq.~\eqref{eq:fermi-stats_term}, the factor
$\prod_{i<j}\sin\left[(\pi/L)(x_i-x_j)\right]$ 
enforces antisymmetry under particle exchange and imposes the correct nodal structure. Similar trial wave functions have been used in other QMC works, see, for example, Refs.~\cite{Astrakharchik2007, Macia2014}. The accuracy of the product ansatz is validated by comparison with DMC results (Appendix~\ref{sec:DMC_study}), which are exact for bosons within statistical uncertainties and subject only to the fixed-node approximation for fermions.

\subsection{Quantum Monte Carlo methods}
To study the ground state properties of the system, we employ QMC techniques, which provide accurate non-perturbative estimates for strongly correlated systems. Specifically, we use the VMC method as the main tool for computing energetic and structural observables, and the Diffusion Monte Carlo (DMC) method for benchmarking our trial wave function in selected characteristic cases. These benchmarking results are presented in Appendix~\ref{sec:DMC_study}.

\subsubsection{Observables}

The main observable calculated in the many-body case is the ground state energy. In addition, we compute several observables to describe spatial correlations in the system.

We calculated the density $n(x)$ to quantify how particles are distributed within the box
\begin{align}
\label{eq:density_profile}
n(x) =  \left \langle \sum_{i=1}^{N} \delta(x-x_i) \right\rangle
\end{align}
where brackets $\langle\cdots\rangle$ denote the expectation value over the ground state distribution.

We also calculate the pair distribution function $g^{(2)}(x, x')$, which measures the probability of finding a pair of particles, one occupying $x$ and the other $x'$, or vice versa, defined as
\begin{align}
g^{(2)}(x, x') \! = \! 2 \! \left\langle \sum_{i < j} \delta(x - x_i)\delta(x' - x_j) \right\rangle\!.
\end{align}
Due to the absence of translational invariance, $g^{(2)}(x,x')$ depends on two positions $(x,x')$ and cannot be reduced to a function depending only on the relative positions $(x-x')$. This function captures short- and long-range spatial correlations.

Finally, we calculate the Leggett bound~\cite{leggett_superfluid_1998} to quantify superfluid response. The Leggett bound provides an upper limit to the superfluid fraction,
\begin{align}
\label{eq:legget_bound}
\frac{\rho_S}{\rho} \leq \frac{1}{\expval{n(x)} \expval*{\frac{1}{n(x)}}},
\end{align}
where $n(x)$ is the density profile defined in Eq.~\eqref{eq:density_profile}.
These observables together allow us to interpret the energy results and identify signatures of localization, clustering, or ordering driven by the interaction parameters.

All Monte Carlo results carry statistical uncertainties inherent to the sampling process. These are estimated following the error bunching method explained in more detail in Ref.~\cite{Flyvbjerg_Petersen_1989}, but are typically smaller than the symbol size and thus not visible in the plots we show in this work.

\section{Two Particle Analysis}
\label{sec:two_particle}

\begin{figure}[t]
\centering
\includegraphics[width=\linewidth]{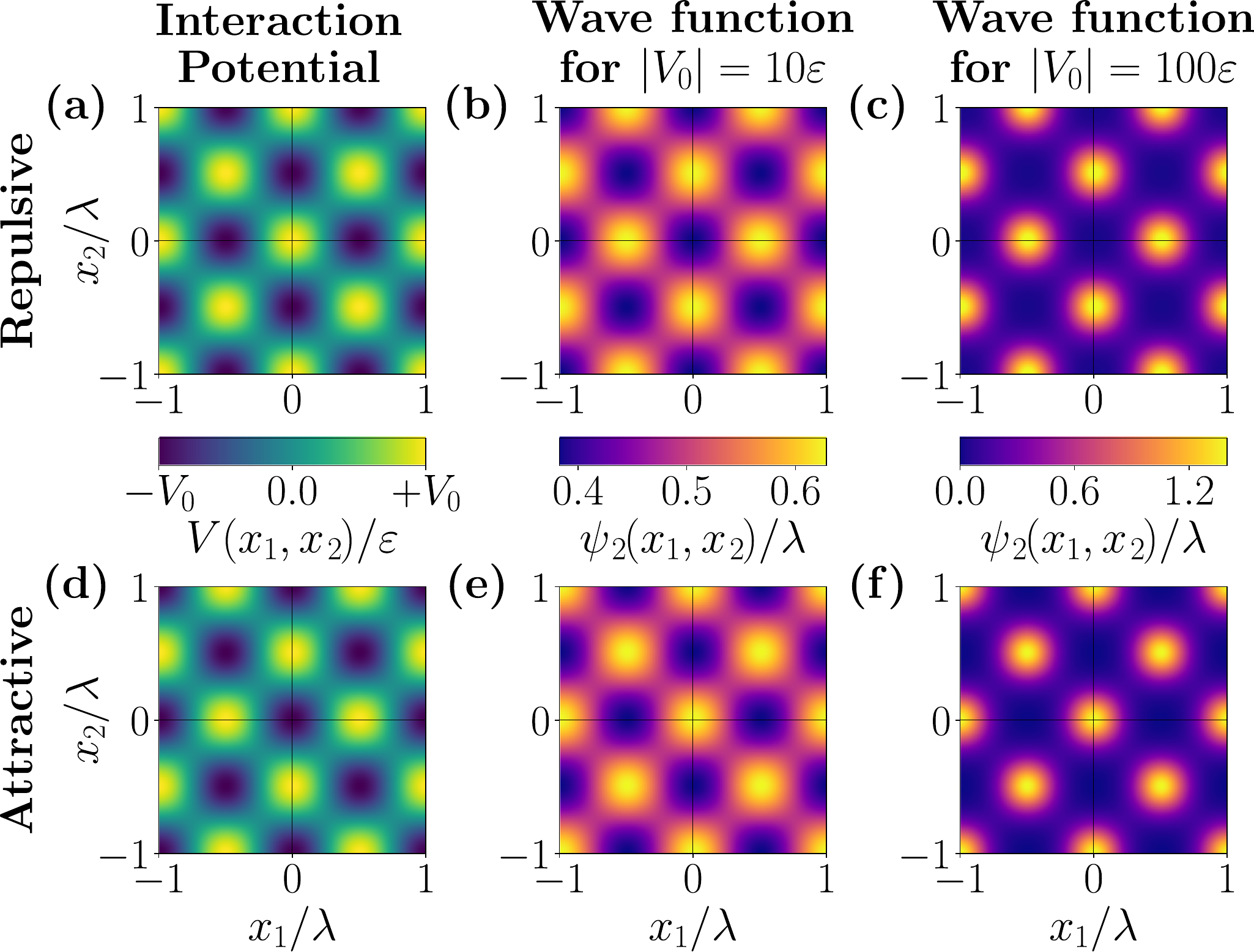}
\caption{Long-range interaction potential and two-body wave functions. Panels (a) and (d): cavity-mediated long-range interaction given by the second term of Eq.~\eqref{eq:1D_hamiltonian_dimensionless} for the repulsive and attractive cases, respectively. Panels (b) and (c): ground state two-body wave function $\psi(x_1, x_2)$ for repulsive interactions, obtained via imaginary time evolution, at interaction strengths $V_0 = 10\varepsilon$ and $V_0 = 100\varepsilon$. Panels (e) and (f): ground state two-body wave functions for the corresponding attractive case.
}
\label{fig:long_range_interaction}
\end{figure}

Now that we have defined the full Hamiltonian and the numerical methods used throughout this study, we pass to the actual analysis of the system. We choose, as a starting point, the simplest case: two particles interacting only via long-range interactions and confined to a box with periodic boundary conditions. In this case, we use the Imaginary Time Evolution (ITE) scheme (defined in Sec.~\ref{sec:two_body_wf_ITE}) to find the ground state properties exactly.

Although seemingly trivial, the two-particle case is still interesting for several reasons. On one hand, it represents the minimal setting in which long-range interaction appears, and thus, interesting physical behaviors emerge for the first time. On the other hand, the wave function of the two-particle solution is used in the construction of the wave function in the many-particle case.

\subsection{Wave function structure and symmetries}

To study how the interactions affect the wave function structure, we show in Fig.~\ref{fig:long_range_interaction} the interaction potential and the resulting two-body wave function for characteristic cases with $|V_0| = 10\varepsilon$ and $|V_0| = 100\varepsilon$ in both the attractive, $V_0 < 0$, and repulsive, $V_0 > 0$, cases.

We find that a characteristic checkerboard pattern emerges. Remarkably, this pattern in the wave function is opposite to the pattern in the interactions; that is, maxima in the interaction correspond to minima in the wave function, and vice versa. This result reflects the tendency of the system to minimize its energy by occupying the regions where the interaction between particles is minimal.

In addition, an important feature that already emerges in the two-particle case 
is the symmetry relating repulsive and attractive interactions. By comparing Fig.~\ref{fig:long_range_interaction}~(b) with (e) and Fig.~\ref{fig:long_range_interaction}~(c) with (f), one observes that the wave functions exhibit identical patterns, but shifted by half a period, following the property of the potential,
\begin{align}
V(x_1,x_2)  =  -V \! \left(x_1 + \frac{\lambda}{2}, x_2\right) = -V \! \left(x_1 , x_2+ \frac{\lambda}{2}\right).
\end{align}

In the attractive case, the diagonal $(x_1=x_2)$ terms of the wave function are strongly populated, indicating that the two particles tend to occupy the same position, similarly to what happens for attractive short-range interactions. On the contrary, for long-range interactions, the shifted diagonals $(x_1=x_2+\ell\cdot\lambda$, with $\ell=\pm1, \pm2, ...)$ are as well populated with the same weight.
Physically, the interaction potential is attractive in regions where the product of the cosine functions is positive, see the second term of  Eq.~\eqref{eq:1D_hamiltonian_dimensionless}. In contrast, in the repulsive case, the wave function is suppressed along the diagonal as the particles avoid each other and preferably occupy positions separated by half a period, where the product of cosines in the second term of Eq.~\eqref{eq:1D_hamiltonian_dimensionless} is negative. The way the system lowers its energy depends on the sign of the interaction, which in turn causes a shift in the characteristic pattern of the wave function.

To show how the amplitude $V_0$ of the long-range interactions affects the wave function, we show in panels~(b) and~(c) of Fig.~\ref{fig:long_range_interaction} the wave function for $|V_0|=10 \, \varepsilon$ and $100 \, \varepsilon$. It becomes evident that higher interaction amplitudes lead to stronger localization of the particles within the interaction wells, both in the attractive and repulsive cases. In the limit of large $V_0$, the wave function fragments and resembles the ground state of a harmonic oscillator repeated periodically across a grid of wells, following the structure of the cavity-mediated potential.

\section{Ground state properties of an ideal Bose gas with cavity-mediated interactions}
\label{sec:many_body_bosons}

\begin{figure}[t]
\centering
\includegraphics[width=\linewidth]{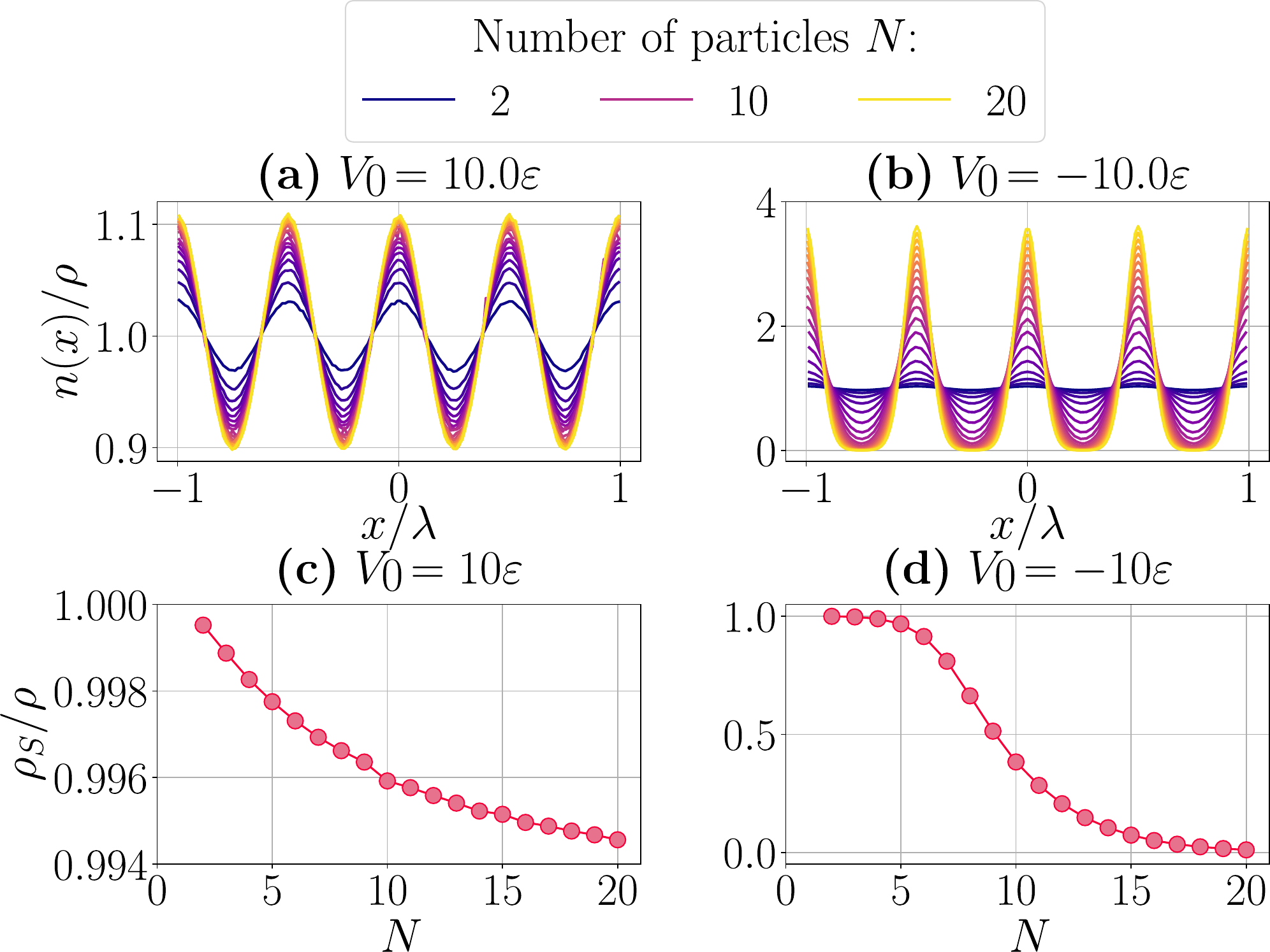}
\caption{
Density profiles $n(x)/\rho$ for various numbers of particles $N = 2, 3, \dots, 20$ obtained by Metropolis sampling, and the corresponding superfluid fraction estimated by the Leggett bound~\eqref{eq:legget_bound}. 
The top row, panels (a) and (b), shows the density profiles $n(x)/\rho$, while the bottom row, panels (c) and (d), displays the superfluid fraction $\rho_S/\rho$. 
The amplitude of $n(x)/\rho$ increases with the number of particles $N$. Larger amplitude corresponds to systems with more particles, both for repulsive ($V_0 > 0$) and attractive ($V_0 < 0$) interactions.
Panels (a) and (c): repulsive interactions with $V_0 = 10\varepsilon$. 
Panels (b) and (d): attractive interactions with $V_0 = -10\varepsilon$. 
All simulations are performed by fixing the length of the system to $L = 2\lambda$ and changing the density as $\rho=N/L$.
}
\label{fig:density_profiles_leggett_bound}
\end{figure}

Using the VMC method, we analyze the ground state properties of an ideal Bose gas subject exclusively to cavity-mediated long-range interactions, as described by the Hamiltonian Eq.~\eqref{eq:1D_hamiltonian_dimensionless} in the absence of short-range interactions, with $g = 0$. This simplified setting allows us to isolate the effects of the periodic, infinite-range interaction and examine its impact on observables such as the density profile, superfluid response, and energy scaling.

\subsection{Density profile and superfluid response}
\label{sec:density_profile}

The density profiles obtained from our Monte Carlo simulations are shown in panels Fig.~\ref{fig:density_profiles_leggett_bound}~(a) and (b) and display periodic modulation for both attractive ($V_0 < 0$) and repulsive ($V_0 > 0$) interactions. These modulations reflect the underlying periodicity of the cavity-mediated interaction, which explicitly breaks the translational invariance of the non-interacting limit ($V_0 = 0$) of the model Hamiltonian, Eq.~\eqref{eq:1D_hamiltonian_dimensionless}. In experimental realizations, continuous translational invariance is already externally broken by the cavity mirrors and the standing-wave mode structure, leaving only a discrete translational symmetry even at $V_0 = 0$~\cite{baumann_dicke_2010, Baumann2011}.

For a larger number of particles, the density peaks become higher, signaling a stronger apparent localization, for both attractive and repulsive interactions. Nevertheless, the effect is significantly stronger for attractive interactions. In the repulsive case, the valleys between peaks remain finite, suggesting that particles stay delocalized across the system, eventually resulting in a coherent system. In contrast, in the attractive case, the minimal density rapidly approaches zero, indicating a tendency for the system to fragment and lose coherence.

Remarkably, attraction has different consequences depending on the range of interaction. Bosons with short-range interactions described by scattering length $a$ form bright solitons~\cite{McGuire64}, where all particles form a deeply bound state of size $\xi \propto a/N$. The resulting density at the center-of-mass position of the soliton rapidly grows as the number of particles is increased, $n(0)\propto N/\xi \propto N^2/a$. Strictly speaking, the average density is constant due to translational invariance, as the soliton can be located at any point. Adding another boson to the existing $N$ leaves this picture unchanged: the new particle is found in the vicinity of the soliton with very high probability.

For long-range interactions, the situation is quite different. First, translational invariance is broken, so the density depends on the position. Second, the system supports a series of localized states rather than a single bound state, consistent with the periodicity of the interaction potential. Finally, adding a particle does not favor any particular site: even if all $N$ particles were occupying the same minimum of the interaction potential, the new particle is equally likely to be found at that minimum or at any other.

The result, particularly for the attractive regime, is a clustered, self-organized phase characterized by pronounced periodic density modulations. Throughout this paper, we refer to this phase as a \emph{self-organized state}.

Focusing on the repulsive regime ($V_0>0$), we find that, for finite particle numbers $N$, the density profiles at fixed system size $L=2\lambda$ exhibit weak periodic modulations. 
However, this modulation is a finite-size artifact of the fixed-box geometry and does not signal an ordering transition. In the thermodynamic limit, repulsive cavity-mediated interactions are known to suppress density fluctuations at the cavity wavevector rather than enhance them~\cite{mottlRotontypeModeSoftening2012, baumann_dicke_2010}. We confirm this directly in the thermodynamic-limit analysis presented in Sec.~\ref{sec:kac-scaling}.

Given the presence of periodic modulation, one might expect suppressed coherence in the system. However, the persistence of coherence, particularly in the repulsive regime, suggests the possible presence of superfluidity. To quantify it, we compute the superfluid fraction with Eq.~\eqref{eq:legget_bound} using the density profiles shown in Fig.~\ref{fig:density_profiles_leggett_bound}~(a) and (b).

The superfluid response is very different in the attractive and repulsive cases [note the different vertical scales in Fig.~\ref{fig:density_profiles_leggett_bound}~(c) and (d)]. Indeed, in the attractive regime, the superfluid fraction rapidly decays with increasing $N$, consistent with the idea that particles localize around the potential minima and lose phase coherence. On the other hand, the repulsive regime retains a finite superfluid fraction for large $N$, consistent with a delocalized phase.

\subsection{Energy scaling and extensivity}
\label{sec:energy_extensive}

\begin{figure}[t]
\centering
\includegraphics[width=\columnwidth]{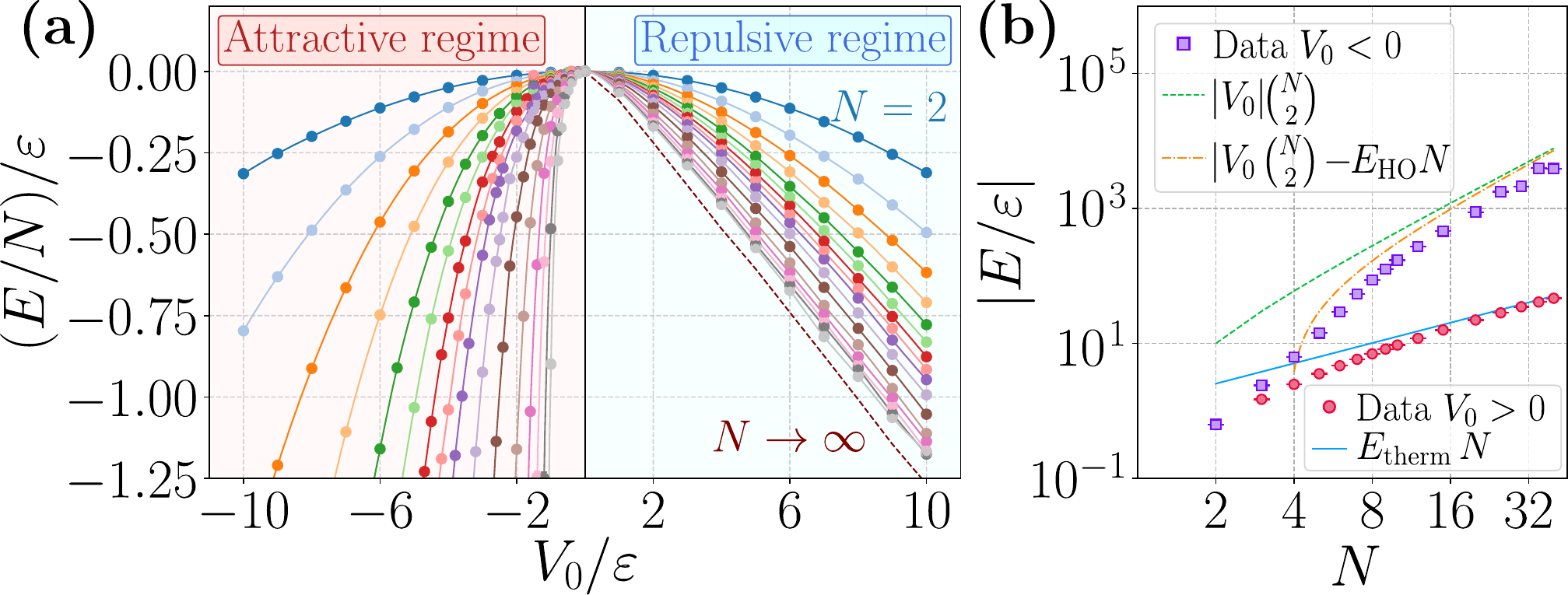}
\caption{
ground state energy of bosons interacting via a cavity-mediated long-range potential. Symbols show VMC results. Panel (a): energy per particle $E/N$ as a function of interaction strength $V_0$ for various particle numbers $N = \left\{2, 3, 4, 5, 6, 7, 8, 9, 10, 12, 15, 20, 25, 30, 35, 40\right\}$. In the repulsive case ($V_0 > 0$), an extrapolated estimate of the thermodynamic limit is shown as a dashed maroon line. In contrast, the attractive case ($V_0 < 0$) leads to a divergent energy. Panel (b): Absolute value of the energy, $|E|$, for the specific case of $|V_0|=10\varepsilon$ as a function of the number of particles, $N$, for both interaction signs, compared to the asymptotic scaling law, Eq.~\eqref{eq:scaling_attractive}, for the attractive case, and $E_{th} N$ for the repulsive one.
All simulations are performed at constant density $\rho = 1 \lambda^{-1}$ and changing the length $L=N/\rho$.
}
\label{fig:energy_only_LR}
\end{figure}

To characterize the finite-size scaling of the energy at a fixed amplitude of interactions $V_0$, we study the energy dependence on the number of particles while keeping the density fixed at $\rho = 1\lambda^{-1}$. Figure~\ref{fig:energy_only_LR}~(a) reports the energy per particle $E/N$ as a function of $V_0$, while Fig.~\ref{fig:energy_only_LR}~(b) shows the total energy $E(N)$ as a function of the number of particles for the amplitude of interactions fixed to $|V_0| = 10\varepsilon$.

In the absence of long-range interaction, $V_0 = 0$, we recover the trivial result for a homogeneous ideal Bose gas,
\begin{align}
E = 0.
\end{align}
As the long-range interaction is introduced, the energy, $E$, becomes negative in both the attractive and repulsive regimes, reflecting the formation of correlated states in regions where the interaction potential is negative. While this lowering of energy is often associated with spatial clustering, in this case, the particles remain delocalized and distribute themselves across the periodic minima of the cavity-mediated potential.

For $N=2$ particles, the energy decreases as the amplitude of the interaction $|V_0|$ is increased; however, the resulting energy is exactly the same for repulsive and attractive interactions of the same strengths, i.e., for amplitudes $\pm V_0$. 
This property was anticipated in Sec.~\ref{sec:two_particle}, where we discussed that in the two-particle case both solutions have similar wave functions, but displaced in space, and share the same energy.
Instead, by increasing further the number of particles, the system properties become drastically different between repulsive and attractive cases.

In the repulsive regime, $V_0 > 0$, we observe that in the limit of a large number of particles, the energy per particle converges toward a constant value,
\begin{align}
\lim_{N \to \infty} \frac{E(N)}{N} = \text{const},
\end{align}
indicating that the system energy is extensive and possesses a well-defined thermodynamic limit. To extract the thermodynamic energy, we perform a linear extrapolation of $E/N$ versus $1/N$ for each value of $V_0$, with the intercept at $1/N = 0$ corresponding to the infinite-particle limit. The extrapolation to the thermodynamic limit is shown in Fig.~\ref{fig:energy_only_LR}~(a) as the dashed maroon line.

In Fig.~\ref{fig:energy_only_LR}~(b) we show the dependence of the absolute value of the ground state energy on the number of particles in two characteristic cases, $V_0=10\varepsilon$ (repulsion) and $V_0=-10\varepsilon$ (attraction). In the repulsive case, the energy grows linearly with particle number, consistent with extensive behavior. This confirms the stability of the gas state and supports the thermodynamic analysis above.

In contrast, the attractive regime exhibits markedly different behavior. As seen in Fig.~\ref{fig:energy_only_LR}~(a) and (b), the energy per particle continues to decrease without any saturation
\begin{align}
\frac{E(N)}{N} \to -\infty,
\end{align}
and no thermodynamic limit is reached at fixed $V_0$. This behavior is characteristic of collapse: particles bunch together into the periodic minima of the interaction potential, and the interaction energy dominates. In the large-$N$ limit, the particles are highly localized in the vicinity of the minima, and the ground state energy approaches the zero-temperature energy of a classical system, given by
\begin{align}
\label{eq:scaling_attractive}
E_{\text{int}}^{\text{clas}}(N)= -|V_0|\frac{N(N-1)}{2}.
\end{align}

Nonetheless, the total energy remains less negative than this pure interaction scaling, indicating that kinetic energy still plays a non-negligible role in moderating the collapse. While in a classical system at zero temperature, the particles freeze in the minimum of the potential energy, a quantum system cannot do that due to the Heisenberg uncertainty principle. Quantum fluctuations lead to a finite kinetic energy and an additional increase of the potential energy. The ground state energy of a harmonic oscillator can approximate the first correction due to the quantum fluctuations,
\begin{align}
E=-|V_0| \frac{N(N-1)}{2} + N \frac{\hbar\omega_0}{2} + \cdots,
\end{align}
where the level spacing $\omega_0$ is obtained by the harmonic approximation of the minimum of the interaction potential:
\begin{align}
\label{eq:approximation_V}
V_0 \cos^2\left(\frac{2\pi}{\lambda}x\right) &\approx V_0 \left(\frac{2\pi}{\lambda}\right)^2 \left( x-\frac{\lambda}{4} \right) ^2 \nonumber \\ &=\frac{1}{2}m\omega^2_0 \left( x - \frac{\lambda}{4}\right)^2 \equiv E_{\text{HO}}.
\end{align}

This correction is shown in Fig.~\ref{fig:energy_only_LR}~(b) as the orange dashed line. We observe a slight improvement, as the resulting energy is lower than the previous approximation. However, higher-order corrections should be considered, since the behavior is still not fully captured.

Together, the results in Fig.~\ref{fig:energy_only_LR}~(a) and (b) highlight a clear contrast at fixed $V_0$: the repulsive regime remains stable and extensive, while the attractive regime exhibits super-extensive scaling and collapse. This behavior is a direct consequence of the all-to-all character of the cavity-mediated interaction potential. 

Although any pairwise Hamiltonian contains $N(N-1)/2$ interaction terms, for short-range interactions the contributions from distant particles become negligible, so that the interaction energy grows linearly with $N$ and the energy per particle converges in the thermodynamic limit. In contrast, the cavity-mediated interaction is non-additive, as each of the $N(N-1)/2$ particle pairs provides a significant contribution to the total energy, causing the interaction energy to scale as $N^2$ rather than $N$~\cite{Defenu2023}. Consequently, observing extensive scaling at fixed interaction amplitude $V_0$ is not sufficient to establish the thermodynamic limit of the long-range system. We therefore reexamine both regimes in Sec.~\ref{sec:kac-scaling} using the Kac-rescaled interaction $V_0\to V_0/N$.

\section{Effects of short-range interactions on cavity-mediated systems}

\begin{figure}[t]
\centering
\includegraphics[width=\linewidth]{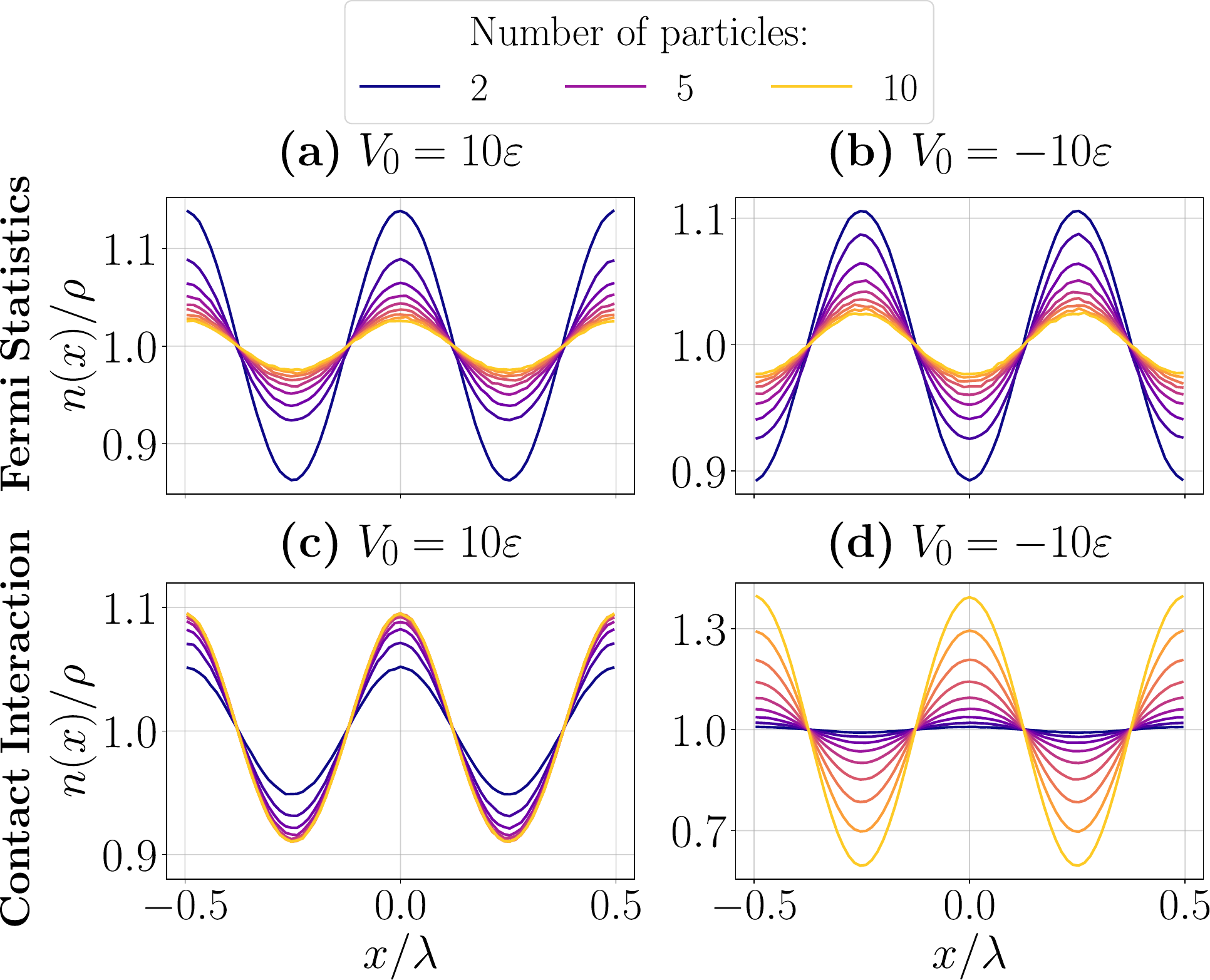}
\caption{
Density profiles $n(x)/\rho$ in the presence of short-range correlations for particle numbers $N=\{2,3, 4, 5, 6, 7, 8, 9, 10\}$.
Panel (a): fermionic system with repulsive long-range interaction ($V_0 > 0$).
Panel (b): fermionic system with attractive long-range interaction ($V_0 < 0$).
Panel (c): bosonic system with contact interaction $\gamma > 0$ and repulsive long-range interaction.
Panel (d): bosonic system with contact interaction and attractive long-range interaction.
All simulations are performed by fixing the length of the system $L = \lambda$ and changing the density $\rho=N/L$.
}
\label{fig:density_contact_fermi}
\end{figure}

\label{sec:short_range_and_fermi}
To study the effects of cavity-mediated long-range interactions, we explore quantum many-body systems incorporating either short-range contact interactions or fermionic statistics. Using the VMC method, we calculate observables such as the ground state energy, density profiles, pair correlation functions, and the static structure factor. These quantities allow us to characterize how the interplay between short-range and cavity-mediated long-range interactions modifies the physical behavior and stability of the system.

To establish benchmarks, we consider two well-understood limits in the absence of cavity-mediated interactions: a weakly interacting Bose gas with short-range repulsion and an ideal Fermi gas governed solely by quantum statistics. The Lieb--Liniger model accurately describes the weakly interacting Bose gas, whose exact solution through the Bethe ansatz provides access to ground state energies and correlations at all interaction strengths.

Properties of a Bose system in the absence of long-range interactions are governed by the dimensionless Lieb–Liniger parameter $\gamma = 2/\rho |a_{\text{1D}}|$, where $\rho = N/L$ denotes the average particle density. The parameter $\gamma$ controls the crossover from weak ($\gamma \ll 1$) to strong repulsion ($\gamma \gg 1$); the former corresponds to the Gross–Pitaevskii regime, while the latter approaches the Tonks–Girardeau limit of impenetrable bosons. Our simulations focus on a weakly interacting regime with a coupling strength $g = -\lambda/a_{\text{1D}} = 1.0 \, \varepsilon \lambda$, corresponding to $\gamma = 2$ at unit density ($\rho = 1\lambda^{-1}$).

On the other hand, the ideal Fermi gas represents the non-interacting limit of a fermionic system, where the Pauli exclusion principle effectively acts as a repulsive interaction. At low energies, both bosonic and fermionic systems are encompassed by the universal framework of Luttinger liquid theory~\cite{haldane_luttinger_1981, cazalilla_one_2011, cazalilla_bosonizing_2004}, which captures common properties such as phononic excitations and algebraically decaying correlations, but with different Luttinger parameters.

By comparing these well-established cases with our full model, where both short-range and cavity-mediated infinite-range interactions are present, we gain insight into how the periodic long-range potential alters fundamental properties such as the superfluid response and the structure factor. This comparative analysis helps us interpret deviations from ideal behavior in terms of emergent correlations and the competition between interaction scales, providing a deeper understanding of the stability and phases of cavity-mediated quantum systems.

\subsection{Density profile and frustrated ordering}
\label{sec:frustrated_regime}

\begin{figure}[t]
\centering
\includegraphics[width=\linewidth]{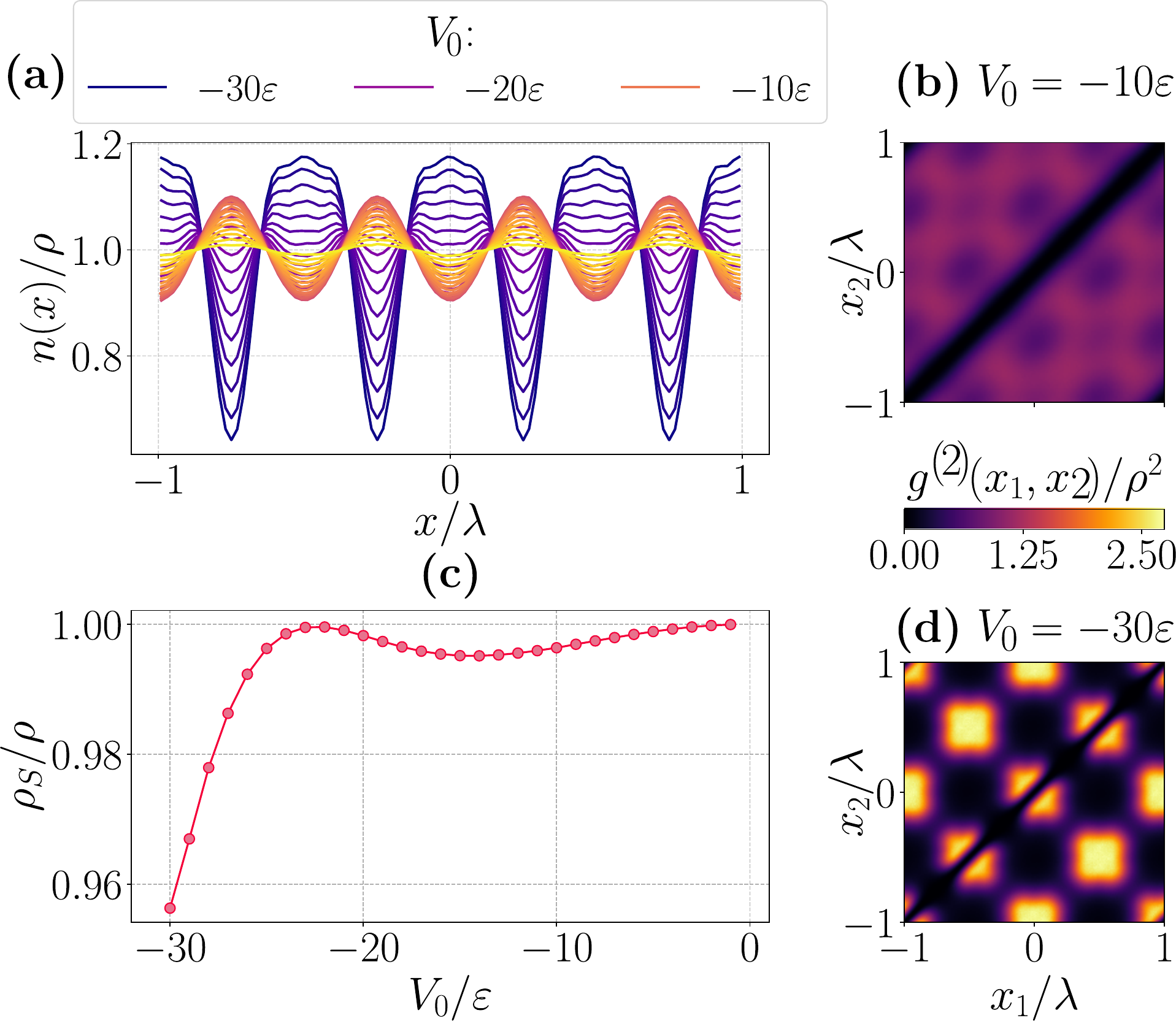}
\caption{
Results for a spinless fermionic system with $N = 5$ particles and attractive cavity-mediated interactions.
Panel (a): density profile $n(x)/\rho$ for varying $V_0$, exhibiting a shift at weak coupling and recovering the bosonic periodic structure at strong coupling.
Panels (b) and (d): two-body correlation function $g^{(2)}(x_1, x_2)/\rho^2$ for $V_0 = -10.0\varepsilon$ and $V_0 = -30.0\varepsilon$, respectively.
Panel (c): superfluid fraction obtained from the Leggett bound, Eq.~\eqref{eq:legget_bound}, showing a crossover in the superfluid response.
All simulations are performed by fixing the system size $L = 2\lambda$ and density $\rho=2.5 \,\lambda^{-1}$.
}
\label{fig:fs_regime_frustrated}
\end{figure}

We begin by analyzing the density profiles $n(x)/\rho$ and two-body correlation functions $g^{(2)}(x_1,x_2)/\rho^2$ in the presence of short-range repulsion, either from contact interaction or Fermi statistics. In the repulsive regime ($V_0 > 0$), both the cavity-mediated and short-range interactions discourage particles from occupying nearby positions. As expected, the density profiles remain relatively flat, with weak periodic modulations slightly softened by repulsion.

In Fig.~\ref{fig:density_contact_fermi} we show the density profile both for a system with contact interaction and a fermionic system. It is remarkable to see how, in the fermionic case, the density profile is not as directly analogous to the purely long-range interaction case as it is for the contact interaction case. Although both fermions and the cavity-mediated interaction introduce a repulsive type of constraint, there is a competition; the system tends to behave more gas-like and lose the periodicity enforced by the cavity potential. This effect is more noticeable in the attractive regime (discussed further below), but it comes from the same fundamental mechanism.

For instance, when $V_0 = -10\varepsilon$, the cavity-mediated interaction becomes less dominant as the particle number increases, allowing the fermionic repulsion to push the system toward a Fermi-gas-like configuration, though still retaining some imprint of the periodic potential. In the bosonic case with contact interaction, a similar effect could arise; however, the contact repulsion is much weaker than the cavity interaction, so the system behaves nearly identically to the ideal Bose gas with only long-range interactions.

The situation becomes more interesting in the attractive regime ($V_0 < 0$), where the cavity-mediated interaction tends to localize particles in its potential minima, while short-range repulsion resists clustering. This competition creates frustration between the two interactions.

It is worth pointing out that frustration only emerges in the many-body setting, and under our assumed system size $L = \lambda$. In larger systems, frustration tends to appear only at higher particle numbers, since the additional space provides more freedom to satisfy interaction constraints. The onset of frustration, therefore, depends not only on the number of particles but also on the system size, which together determine whether all pairwise constraints are simultaneously fulfilled. In the thermodynamic limit, the long-range attraction eventually dominates for any fixed $V_0 < 0$, and this regime vanishes. More generally, we expect that frustration will not emerge for densities $\rho \leq 1$, since the available space per particle is then sufficient to satisfy both the short-range and long-range constraints simultaneously.

To explore this effect in detail, we focus on the fermionic system with $N = 5$ particles—a size large enough to exhibit spatial structure, but small enough to resolve individual correlations. We choose the fermionic case because the effects of frustration are visually more pronounced due to strong Pauli repulsion. Similar effects are also present in bosonic systems with contact interactions, though they are significantly weaker and harder to visualize with the same clarity.

Figure~\ref{fig:fs_regime_frustrated} shows both the density profile and the two-body correlation function for increasing attractive strength. Panel (a) reveals that for moderate attraction ($1 < |V_0|/\varepsilon < 20$), the density stays nearly uniform, though faint modulations manifest due to the underlying cavity-mediated interaction.

This effect becomes even clearer when observing the two-body correlation function $g^{(2)}(x_1,x_2)/\rho^2$ in  Fig.~\ref{fig:fs_regime_frustrated}~(b). At $V_0=-10\varepsilon$, which corresponds to an intermediate attraction, the pair distribution presents a very uniform pattern, but with a weak periodic modulation that has maxima now at the zeros of the interaction instead of the minima, corresponding to a half-period phase shift in the density profile, see Fig.~\ref{fig:fs_regime_frustrated}~(a). The system avoids both the interaction maxima and the diagonal $x_1 = x_2$, which is forbidden by Fermi statistics.

This results in a compromise that balances both types of interaction: the attractive cavity-mediated potential, which favors clustering, and the Pauli exclusion principle, which prevents particle overlap. As a result, the particles tend to organize near the zeros of the interaction potential, showing a nontrivial compromise between these opposing tendencies.

As $|V_0|$ increases further, the periodic structure imposed by the cavity-mediated interaction begins to dominate. The system develops sharp density maxima that match with the minima of the interaction potential, similar to what is observed in the long-range-only case. However, even in this limit, the system continues to avoid the diagonal in $g^{(2)}(x,x')$, preserving Pauli exclusion. The width of this forbidden region narrows as the attraction grows stronger, reflecting increased localization around the cavity minima and the growing dominance of the cavity-mediated interaction. This is illustrated in Fig.~\ref{fig:fs_regime_frustrated}~(a), (b) and (d).

This frustrated regime, in which particles are caught between the ordering tendency of the long-range interaction and the repulsive constraint imposed by Fermi statistics, defines a nontrivial configuration, neither an ideal Fermi gas nor the self-organized state found in the purely long-range Bose case. To quantify and further characterize the transition out of this regime, we compute the Leggett upper bound for the superfluid fraction as a function of interaction strength.

The superfluid fraction is expected to remain finite as long as the system retains a delocalized, gas-like character, and to drop rapidly to zero once the particles begin to localize within the cavity minima. This makes it a useful indicator of the crossover between the frustrated regime and the interaction-dominated, symmetry-broken state.

As shown in Fig.~\ref{fig:fs_regime_frustrated}~(c), the superfluid response remains large in the frustrated regime and becomes strongly suppressed once the long-range attraction becomes dominant.

\subsection{Energy scaling and qualitative phase diagram}
\label{sec:energy_sr_fs}

\begin{figure}[t]
\centering
\includegraphics[width=\linewidth]{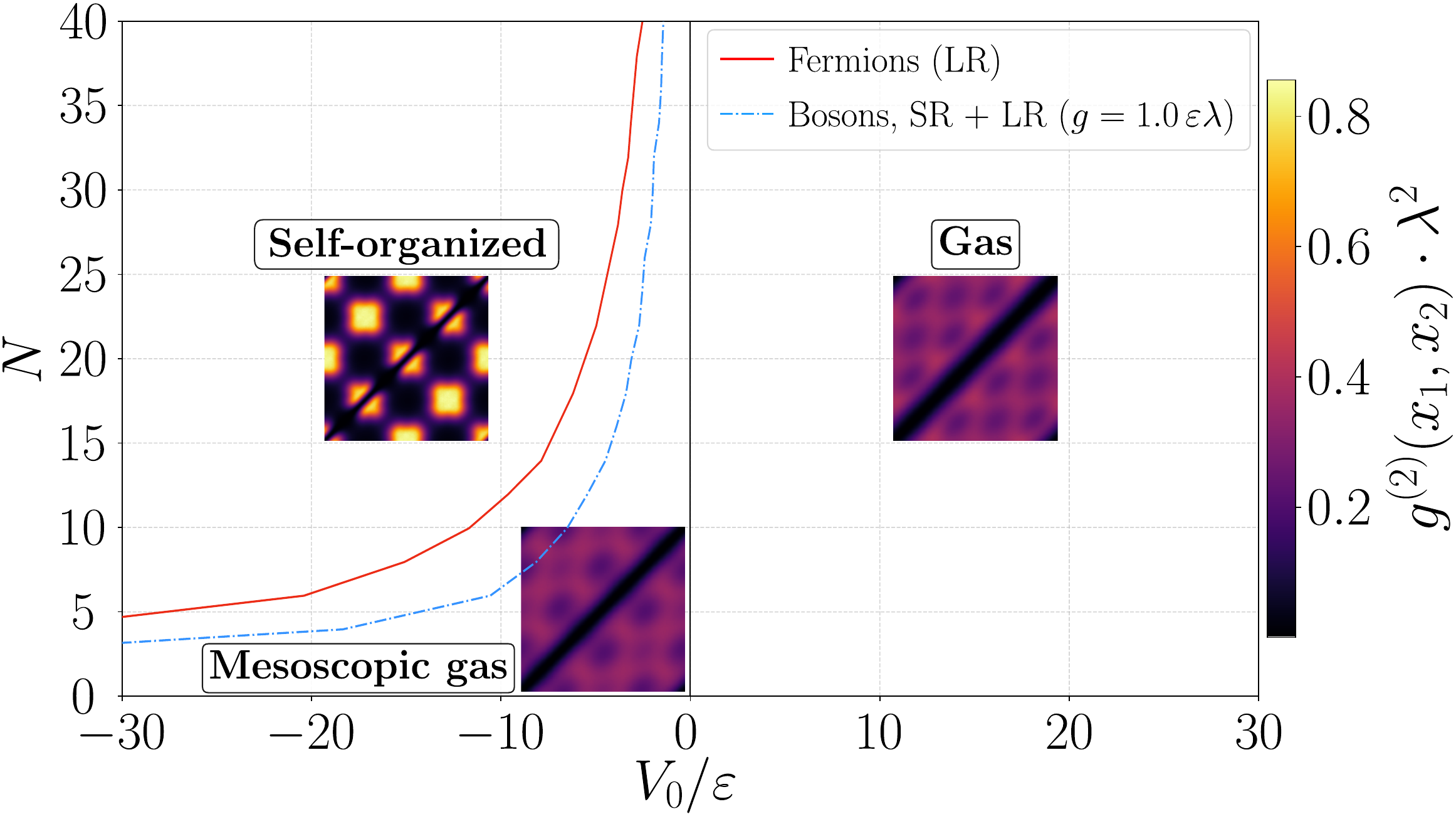}
\caption{
Qualitative phase diagram showing three regimes: self-organized state, mesoscopic gas, and gas phase. The mesoscopic gas is separated from the thermodynamically stable gas phase by the vertical line $V_0 = 0$, and from the self-organized state by the self-organizing onset curve $V_0^c(N)$, obtained from the superfluid fraction as described in the text, and shown as a red line for the fermionic system and as a blue line for bosons with contact interaction ($g = 1.0 \, \varepsilon\lambda$). For purely cavity-mediated long-range interactions, the $V_0 = 0$ line marks the boundary between the gas and self-organized state. Representative pair correlation functions $g^{(2)}(x_1, x_2)/\rho^2$ illustrate the changes in spatial correlations across these regimes; in particular, we use the case of $N = 5$ particles for the fermionic system. Simulations are performed at fixed density $\rho = 2\lambda^{-1}$ with system size $L = N/\rho$, except for the pair correlation functions, which are computed at fixed $L = 2\lambda$ for better visualization.
}
\label{fig:phase_diagram}
\end{figure}

We now extend the energy analysis to systems that include additional short-range repulsion, either through Fermi statistics or weak contact interaction.

For strong long-range interactions, the observed behavior closely resembles that of the purely long-range case. In the repulsive regime, $V_0 > 0$, the energy per particle $E/N$ converges with increasing $N$, confirming the existence of a well-defined thermodynamic limit. In the attractive regime, $V_0 < 0$, the energy decreases with $N$ and shows signs of collapse, although the rate of divergence is reduced due to the presence of repulsive effects.

The most notable difference is found at $V_0 = 0$, where the energy is no longer zero but reflects the baseline contribution from the short-range terms. In the fermionic case, this corresponds to the energy of the ideal Fermi gas, which is significantly higher. In the bosonic case with contact interaction, the shift is much smaller, consistent with the weaker repulsion.

Importantly, the short-range repulsion acts to delay the onset of collapse. This stabilization is especially visible in the fermionic case, where Pauli repulsion remains strong even as $V_0$ becomes increasingly negative. For bosons with contact interaction, the effect is much weaker. For the same $V_0$ as in the fermionic case, it is only noticeable at small $N$, where the relative contribution of the short-range repulsion to the total energy is larger before the long-range attraction dominates. As such, while the overall instability of the attractive regime persists, its onset is significantly modified by the nature and strength of short-range repulsion.

In the previous section, we showed that the emergence of a frustrated regime is reflected in the behavior of the superfluid fraction, as the system is almost fully superfluid in the frustrated regime, $\rho_S/\rho\approx 1$, while the superfluid fraction decreases almost linearly as the system approaches the self-organized state. As a result, the second derivative of $\rho_S/\rho$ is small deep in each regime, $(\rho_S/\rho)''\approx 0$, and its absolute value is maximal in the transition region. 
We define the critical interaction amplitude, $V_0^c(N)$, for which $|(\rho_S/\rho)''|$ is maximal, i.e., the value at which the third derivative vanishes, $(\rho_S/\rho)''' = 0$, restricted to the region where the curve departs from the plateau (see Appendix~\ref{sec:criterion}).

Based on this criterion, we present in Fig.~\ref{fig:phase_diagram} a qualitative phase diagram constructed by identifying, for each number of particles $N$, the critical value of the amplitude of the long-range interactions $V_0^c(N)$.

This phase diagram reveals three distinct regimes. In the repulsive case ($V_0 > 0$), the system remains gas-like across all interaction strengths, exhibiting only a weak periodic modulation induced by the long-range interaction. In the attractive regime ($V_0 < 0$), two qualitatively different phases emerge. For small values of $V_0$, the system remains in a mesoscopic gas phase: the LR attraction is not sufficiently strong to induce localization in the minima of the interaction potential, and the particles remain delocalized, exhibiting gas-like behavior. However, beyond a certain critical value of $V_0$, which depends on the particle number, the system localizes in the minima of the interaction potential, forming what we define as a self-organized state. It is important to point out that the gas-like phase for attractive interactions is a mesoscopic one; that is, it exists only for a finite number of particles and vanishes in the thermodynamic limit.
By increasing the number of particles, the system enters the self-organized phase identified in the long-range-only case and gradually loses all superfluid response.

We also observe that, in the presence of contact interactions, the critical amplitude $V_0$ required for localization is lower than in the fermionic case for the same number of particles $N$. This difference can be attributed to the weaker effective repulsion between bosons with contact interactions, making the system more susceptible to clustering under attractive long-range forces.

The frustrated regime identified here arises from the competition between the cavity-mediated attraction and the short-range or Fermi-statistics constraint at densities above unit filling. 
Its presence is a finite-size effect that vanishes in the thermodynamic limit, as shown explicitly in Sec.~\ref{sec:kac-scaling}, providing a possible explanation for its absence in previous studies focused primarily on the large-particle-number or continuum limit.

\section{Kac prescription and the thermodynamic limit}
\label{sec:kac-scaling}

\begin{figure}[t]
\centering
\includegraphics[width=\linewidth]{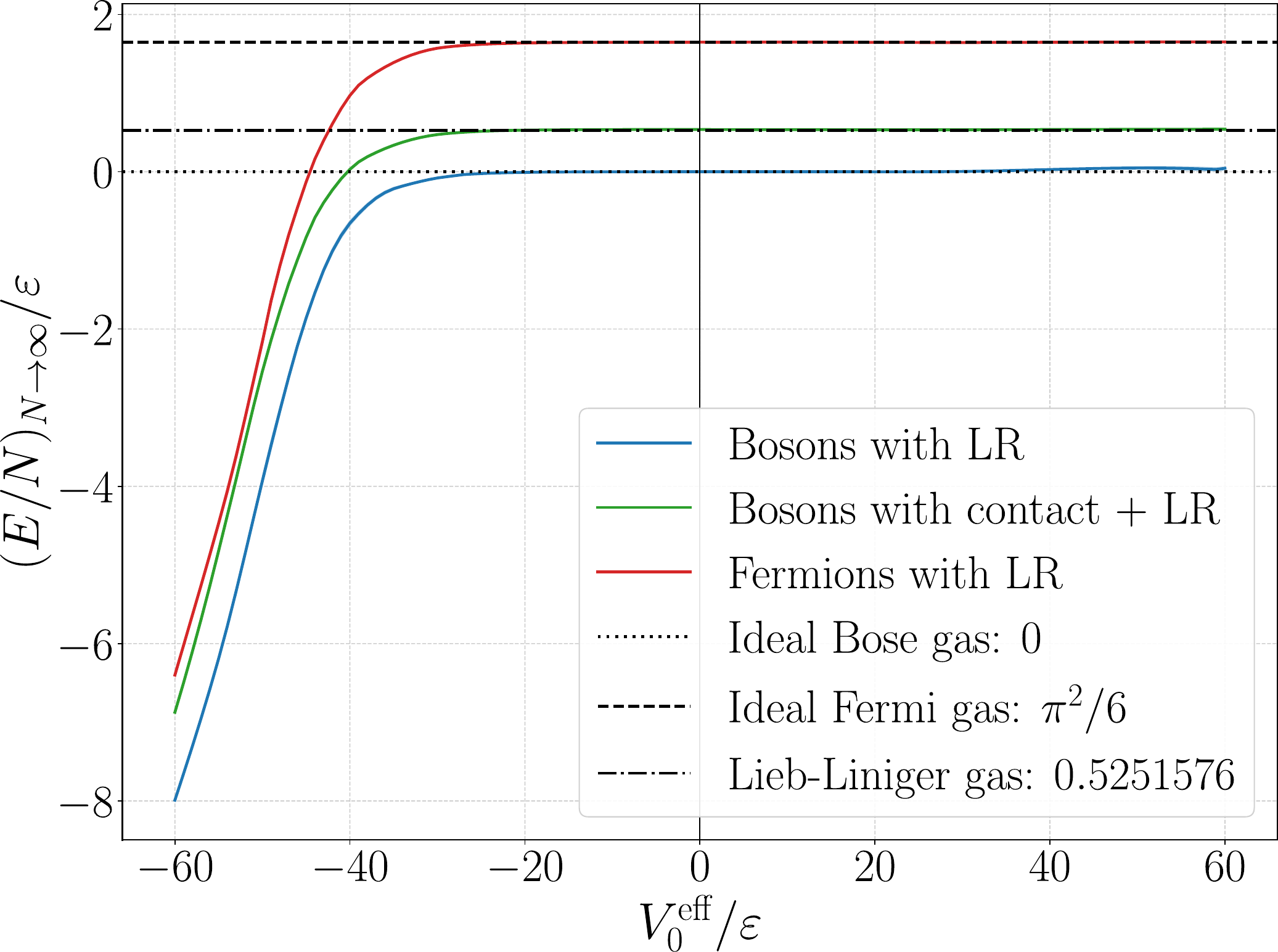}
\caption{ground state energy per particle in the thermodynamic limit for bosons with long-range interactions, bosons with combined short- and long-range interactions, and fermions with long-range interactions, obtained via Kac scaling ($V_0 \to V_0/N$). Statistical uncertainties from the linear extrapolation to $N\to\infty$ are smaller than the linewidth and are not visible on this scale. All simulations are performed at constant density $\rho = 1\lambda^{-1}$, changing the system length as $L=N/\rho$.}
\label{fig:kac_thermo_all}
\end{figure}

In the previous sections, particularly in Secs.~\ref{sec:energy_extensive} and~\ref{sec:energy_sr_fs}, we found that the ground state energy exhibits non-extensive behavior in the presence of strong long-range interactions: the energy per particle fails to converge to a finite value as $N \to \infty$ at fixed $V_0$. This lack of extensivity is a generic feature of all-to-all long-range interactions, for which the interaction energy scales as $N^2$ rather than $N$, owing to the $N(N-1)/2$ contributing pairs~\cite{Defenu2023}. Since we aim to characterize the thermodynamic properties of the system, a suitable procedure is required to obtain a well-defined thermodynamic limit.

This has been addressed in the literature with the so-called Kac prescription. It consists of modifying the Hamiltonian and rescaling the long-range interaction by a prefactor that depends on the total number of particles, ensuring energy extensivity~\cite{Kac1963}. Because this rescaling modifies only the coupling strength and not the interaction range itself, the nonadditive character intrinsic to long-range systems remains intact. At the same time, the energy per particle and other intensive thermodynamic quantities built from it converge to well-defined, $N$-independent values in the thermodynamic limit. For further details, see Refs.~\cite{Kastner2025, Campa2014, Defenu2023}.

In this section, we study both the ground state energy and the density profile in the thermodynamic limit using the Kac prescription.

\subsection{Energy}
\label{sec:kac-energy}

For each fixed value of the effective long-range interaction amplitude, $V_0^{\text{eff}} \equiv NV_0$, we compute the ground state energy per particle $E(N)/N$ at a constant density, $\rho = 1\lambda^{-1}$ (chosen so as to avoid the frustrated regime discussed in Sec.~\ref{sec:frustrated_regime}), for different number of particles $N$, and extrapolate linearly in $1/N$ to zero to obtain the thermodynamic-limit value $(E/N)_{N\to\infty}$ (see Appendix~\ref{sec:extrapolation} for representative examples of this extrapolation). Figure~\ref{fig:kac_thermo_all} shows the thermodynamic energy as a function of the interaction amplitude, $V_0^{\text{eff}}$, for the three cases studied in this work: bosons with long-range interactions only, bosons with combined short- and long-range interactions, and fermions with long-range interactions. In contrast with the fixed long-range interaction amplitude ($V_0=$const) analysis of Secs.~\ref{sec:energy_extensive} and~\ref{sec:energy_sr_fs}, where the attractive regime showed no thermodynamic limit at all, using the Kac prescription, the thermodynamic-limit value is well defined for each value of $V_0^{\text{eff}}$, on both the repulsive and attractive sides.
 
On the repulsive side ($V_0^{\text{eff}} > 0$), the energy in the thermodynamic limit coincides with that of the corresponding non-interacting case for each system: the ideal Bose gas, the Lieb-Liniger gas, and the ideal Fermi gas, respectively. This is consistent with both the experimental and theoretical findings of Ref.~\cite{mottlRotontypeModeSoftening2012}, where no spontaneous ordering is expected for repulsive interactions.
 
On the attractive side, we distinguish two regimes. For weak interaction strength, the long-range interaction is not strong enough to induce the self-organizing phase transition, and the energy remains constant with $V_0^{\text{eff}}$, corresponding to a gas-like phase. Once $|V_0^{\text{eff}}|$ reaches approximately $30\,\varepsilon$, the energy begins to decrease monotonically, indicating the onset of the self-organized phase reported in previous experimental and theoretical studies. We emphasize that, in our results, this onset is not marked by a kink in the energy: the energy is a smooth, monotonic function throughout the entire domain, with no discontinuity in slope. These two regimes are consistent with those found in Refs.~\cite{mottlRotontypeModeSoftening2012, baumann_dicke_2010}, where, depending on the value of $V_0$ (itself set by $\eta$ and $\Delta_c$), the system is found either in a gas-like phase or, for sufficiently strong coupling, in the self-organized phase.
 
Notably, this threshold is essentially the same for all three cases studied, despite their markedly different energies at $V_0^{\text{eff}}=0$. This reflects the fact that, at the density $\rho = 1\lambda^{-1}$ considered here, there is no genuine competition between the long-range ordering tendency and the short-range or Fermi-statistics constraint: as discussed in Sec.~\ref{sec:frustrated_regime}, at this density the available space is sufficient for both constraints to be satisfied simultaneously, so the onset of the self-organized phase is governed by the long-range interaction alone, independently of the additional short-range physics present. The density profile, discussed below, further supports this picture.

\subsection{Density profile}
\label{sec:density_kac}

\begin{figure}[t]
\centering
\includegraphics[width=\linewidth]{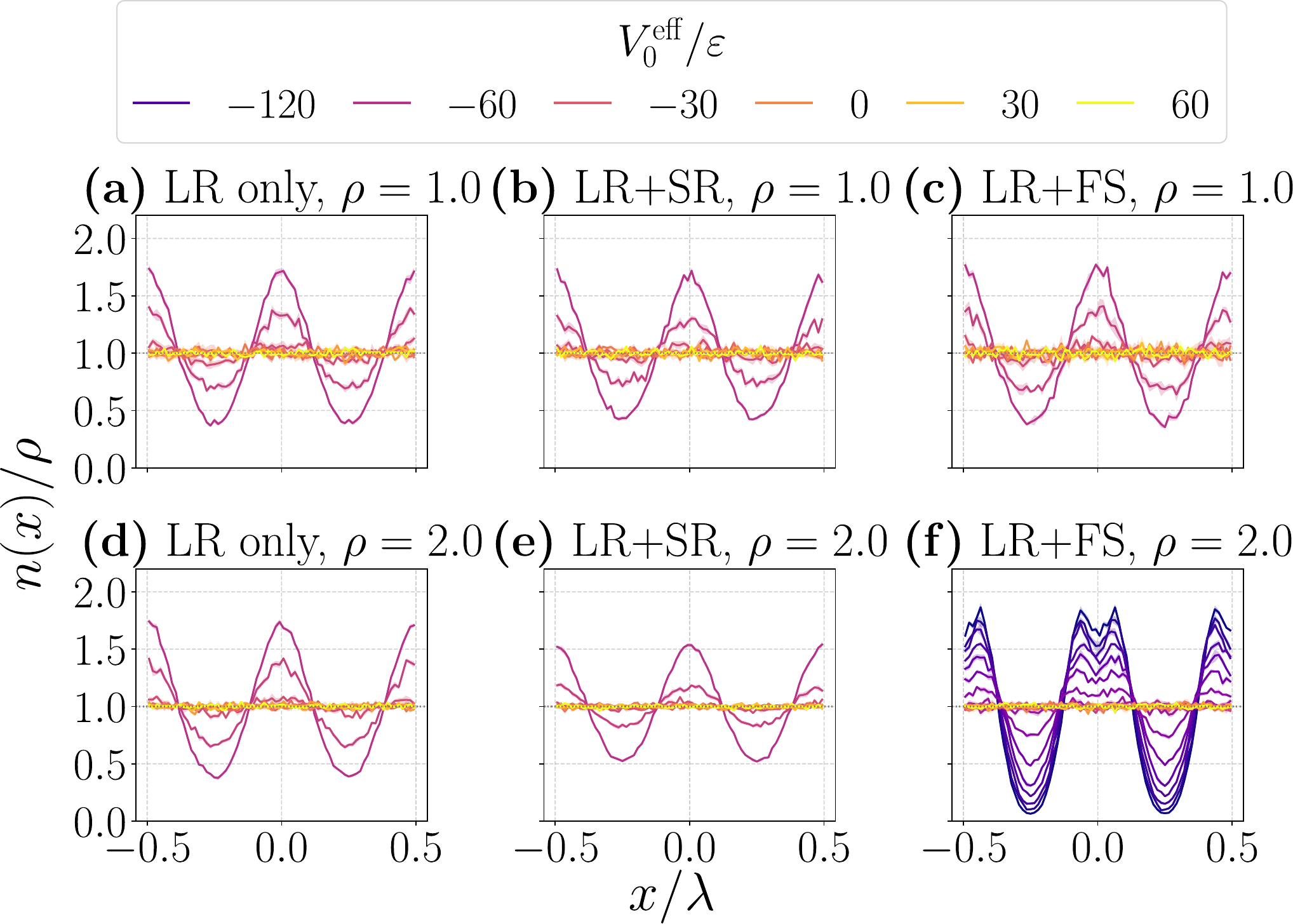}
\caption{Density profiles $n(x)/\rho$ in the thermodynamic limit, obtained by extrapolating $n(x)/\rho$ at fixed $x$ to $1/N \to 0$ along Kac trajectories of constant $V_0^{\text{eff}} = NV_0$, shown for representative values of $V_0^{\text{eff}}/\varepsilon$ common to all panels. Profiles are restricted to the central unit cell, $x/\lambda \in [-0.5, 0.5)$, by translational periodicity. (left column) Bosons with long-range interactions only. (center column) Bosons with combined short- and long-range interactions. (right column) Fermions with long-range interactions. (a)--(c) $\rho = 1\,\lambda^{-1}$; (d)--(f) $\rho = 2\,\lambda^{-1}$. Shaded bands show the statistical uncertainty of the linear extrapolation. All simulations are performed at fixed density $\rho$, with $L = N/\rho$.}
\label{fig:density_profiles}
\end{figure}

To complement the energy analysis, we determine the density profile $n(x)/\rho$ in the thermodynamic limit along fixed Kac trajectories $NV_0 \equiv V_0^{\text{eff}} = \text{const}$. At each spatial point $x$, we extrapolate the density profile linearly in $1/N \to 0$, analogous to the procedure used for the energy in Sec.~\ref{sec:kac-energy}. Owing to the translational periodicity of the system with period $\lambda$, the density profile within a single unit cell fully characterizes the density everywhere. Figure~\ref{fig:density_profiles} shows the resulting thermodynamic-limit density profiles, restricted to a single unit cell, for the three cases studied in this work, at two fixed densities, $\rho = 1\,\lambda^{-1}$ and $\rho = 2\,\lambda^{-1}$, for representative values of $V_0^{\text{eff}}$.
 
In the repulsive regime ($V_0^{\text{eff}} > 0$), the density profile becomes essentially flat, despite some residual noise, as $N \to \infty$ at fixed $V_0^{\text{eff}}$: the periodic modulation observed at finite $N$ in Sec.~\ref{sec:density_profile} vanishes in the thermodynamic limit. This confirms that the finite-$N$ modulation reported earlier was indeed a finite-size artifact and not a genuine ordering instability, consistent with the absence of roton softening in the repulsive regime reported in Refs.~\cite{mottlRotontypeModeSoftening2012, baumann_dicke_2010}.

In the attractive regime ($V_0^{\text{eff}} < 0$), by contrast, the density modulation is progressively enhanced as $|V_0^{\text{eff}}|$ increases, consistent with the self-organized state identified in Sec.~\ref{sec:energy_extensive}: particles remain clustered at the periodic minima of the interaction potential even in the thermodynamic limit. Together with the energy behavior of Sec.~\ref{sec:kac-energy}, this confirms that the qualitative distinction between the repulsive and attractive regimes established in Sec.~\ref{sec:energy_extensive} survives the proper thermodynamic-limit analysis, although the onset of the self-organized state is delayed to larger $|V_0^{\text{eff}}|$ than suggested by the fixed-$V_0$ analysis of Sec.~\ref{sec:energy_extensive}, where ordering appeared to set in for arbitrarily weak attraction. Moreover, the resulting density profile itself remains a well-defined, finite periodic structure in this limit.

We also observe that, for the higher density $\rho = 2\lambda^{-1}$, where frustration was found to occur at finite $N$ in Sec.~\ref{sec:frustrated_regime}, the frustrated phase disappears entirely in the thermodynamic limit. This is consistent with the finite-size origin of frustration discussed in Sec.~\ref{sec:frustrated_regime}: at $\rho = 1\lambda^{-1}$, the available space is sufficient for both the long-range ordering tendency and the short-range or Fermi-statistics constraint to be simultaneously satisfied, so frustration does not arise; only at the higher density $\rho = 2\lambda^{-1}$ does the reduced spacing between particles make the two constraints incompatible at finite $N$. In the thermodynamic limit, however, we instead find only a gas phase, which persists to larger values of $|V_0^{\text{eff}}|$ than at $\rho\;=\;1\lambda^{-1}$, as the competition between the long-range attraction and the short-range or Fermi repulsion makes it harder for the system to collapse into the self-organized phase, delaying its onset to stronger coupling. This gas phase shows none of the peak-shifting behavior characteristic of the finite-size frustrated regime discussed in Sec.~\ref{sec:frustrated_regime}. We note that a similar delay of the self-organization threshold to stronger coupling, in the presence of additional competing physics, is also observed experimentally: comparing the bosonic~\cite{baumann_dicke_2010} and fermionic~\cite{Zhang2021, helsonDensitywaveOrderingUnitary2023} phase diagrams, the fermionic self-organization threshold requires substantially higher pump power (of order $\mathrm{mW}$) than the bosonic case (of order \textmu W), consistent with the general expectation that the fermionic exclusion principle offers greater resistance to the long-range interaction than a weak contact interaction, pushing the self-organization threshold to higher values of $V_0$.

\begin{table*}[t]
\caption{Summary of trial wave functions for different interaction regimes and quantum statistics. 
Here, ``LR'' refers to cavity-mediated long-range interactions (second term of Eq.~\eqref{eq:1D_hamiltonian_dimensionless}), while ``SR'' denotes short-range contact interactions modeled by a delta-function potential. 
Symbols $\text{B}$ and $\text{F}$ indicate bosonic and fermionic symmetry, respectively. 
The factor $\psi_2(x_i,x_j)$ represents the exact two-body wave function for the long-range interaction case. 
Here, $L$ is the size of the system, and $k_c$ is determined by the Bethe-Peierls boundary condition.}
\label{tab:summary_trial_wf}
\centering
\begin{ruledtabular}
\begin{tabular}{ll}
System & Trial wave function \\ 
\hline
Bosons LR &
$\Psi^{\text{(B)}}_{\text{LR}}(x_1, \dotsc, x_N)
 = \prod_{i<j} \psi_2(x_i, x_j)$ \\[4pt]

Bosons SR+LR &
$\Psi^{\text{(B)}}_{\text{SR+LR}}(x_1, \dotsc, x_N)
 = \prod_{i<j} \psi_2(x_i, x_j)
   \prod_{i<j} \cos\!\left[k_c\!\left(|x_i - x_j| - \frac{L}{2}\right)\right]$ \\[4pt]

Fermions LR &
$\Psi_{\text{LR}}^{\text{(F)}}(x_1,\dotsc,x_N)
 = \prod_{i<j} \psi_2(x_i, x_j)
   \prod_{i<j} \sin\!\left[\frac{\pi}{L}(x_i - x_j)\right]$ \\[4pt]

Fermions SR+LR &
No contact interaction can be added because spinless fermions do not interact via $s$-wave scattering \\
\end{tabular}
\end{ruledtabular}
\end{table*}

\section{Conclusions and Outlook}
\label{sec:conclusion}

This paper studies the ground state properties of a one-dimensional quantum system with cavity-mediated long-range interactions in continuous space, using numerical methods across different interaction regimes and quantum statistics. The system displays a wide variety of behaviors as summarized in the phase diagram (Fig.~\ref{fig:phase_diagram}).

We investigate the many-body properties in the absence of short-range interactions, isolating the effects of cavity-mediated long-range potentials at a fixed $V_0$. Such a system exhibits two qualitatively different regimes depending on the sign of the interaction strength. For attractive interactions, particles strongly cluster at the periodic minima of the potential, leading to pronounced density modulations, a vanishing superfluid fraction, and the absence of a well-defined thermodynamic limit; we refer to this regime as a self-organized state. In contrast, repulsive long-range interactions produce a delocalized phase characterized by weak periodic density modulations, a finite superfluid fraction, and a well-defined thermodynamic limit.

The incorporation of short-range repulsion, through either contact interactions or fermionic statistics, significantly modifies the properties of the system. At weak coupling, the cavity-mediated interaction perturbs the density profile only slightly, and the system remains close to its short-range-only counterpart. As attraction increases, the interplay between long-range attraction and short-range repulsion generates a frustrated mesoscopic phase marked by intricate spatial correlations and a finite superfluid fraction. As attraction is increased further, the system undergoes a transition into a self-organized state, reflecting a rich competition between the repulsive forces and clustering tendencies.

Using the Kac prescription, we established the proper thermodynamic limit of the system, addressing the non-extensive scaling found at fixed $V_0$. In the repulsive regime, the thermodynamic-limit energy coincides with that of the corresponding non-interacting system in each case, and the periodic density modulation observed at finite $N$ vanishes, confirming that it is a finite-size artifact rather than a genuine ordering instability. In the attractive regime, a well-defined thermodynamic limit is recovered, and the self-organized state persists. However, its onset is delayed to substantially stronger coupling than suggested by the fixed-$V_0$ analysis. We further find that the frustrated mesoscopic regime identified in the presence of short-range repulsion is itself a finite-size effect: it disappears entirely in the thermodynamic limit, where only the gas and self-organized phases remain.

From a methodological point of view, we developed a many-body trial wave function based on the exact two-body solution of the cavity-mediated potential. This wave function was implemented within both VMC and DMC methods. We validated these methods by benchmarking against known analytical limits (perturbative expansions), exactly solvable cases (ideal Fermi gas, Bethe ansatz for Lieb-Liniger gas), and other numerical methods (imaginary-time propagation for the two-particle case), ensuring consistency across different interaction regimes. Additionally, convergence tests were performed with respect to simulation parameters such as particle number, time step size, and projection time, confirming the robustness and accuracy of our approach. These validations guarantee reliable quantitative predictions for ground state energies and correlation functions in cavity-mediated many-body systems. 

The numerical methods developed here are versatile and can be extended to higher-dimensional or more realistic versions of the cavity-mediated model studied in this work. In particular, one could first consider the limit of a deep transverse lattice ($V_p \to \infty$), freezing motion along $z$ and reducing the system to two dimensions, and then further extend the approach to the full three-dimensional geometry, both corresponding to the actual experimental setup of Ref.~\cite{mottlRotontypeModeSoftening2012}. Additionally, extra terms such as optical lattices, multimode cavity fields, or other position-dependent interactions can be incorporated within the same framework.

In conclusion, this study combines a physically motivated model with flexible numerical tools to explore a range of rich many-body phenomena. It offers both a foundation for future simulations and concrete insight into the behavior of ultracold atoms coupled to cavity fields.

\begin{acknowledgments}
This work has been funded by Grant No. PID2023-147475NB-I00 funded by MICIU/AEI/10.13039/501100011033 and FEDER, UE, by grants 2021SGR01095 from Generalitat de Catalunya, and by 
Project No. CEX2019-000918-M of ICCUB (Unidad de Excelencia María de Maeztu). 
G.E.A. acknowledges the support of the Spanish Ministry of Science and Innovation (MCIN/AEI/10.13039/501100011033, Grant No. PID2023-147469NB-C21), the Generalitat de Catalunya (Grant No. 2021 SGR 01411) and {Barcelona Supercomputing Center MareNostrum} ({FI-2025-1-0020}).
A.R-F. acknowledges additional funding from the Okinawa Institute of Science and Technology Graduate University.
\end{acknowledgments}

\appendix

\section{Comparing energy results using VMC and DMC methods}

\label{sec:DMC_study}

\begin{figure}[t]
    \centering\includegraphics[width=\linewidth]{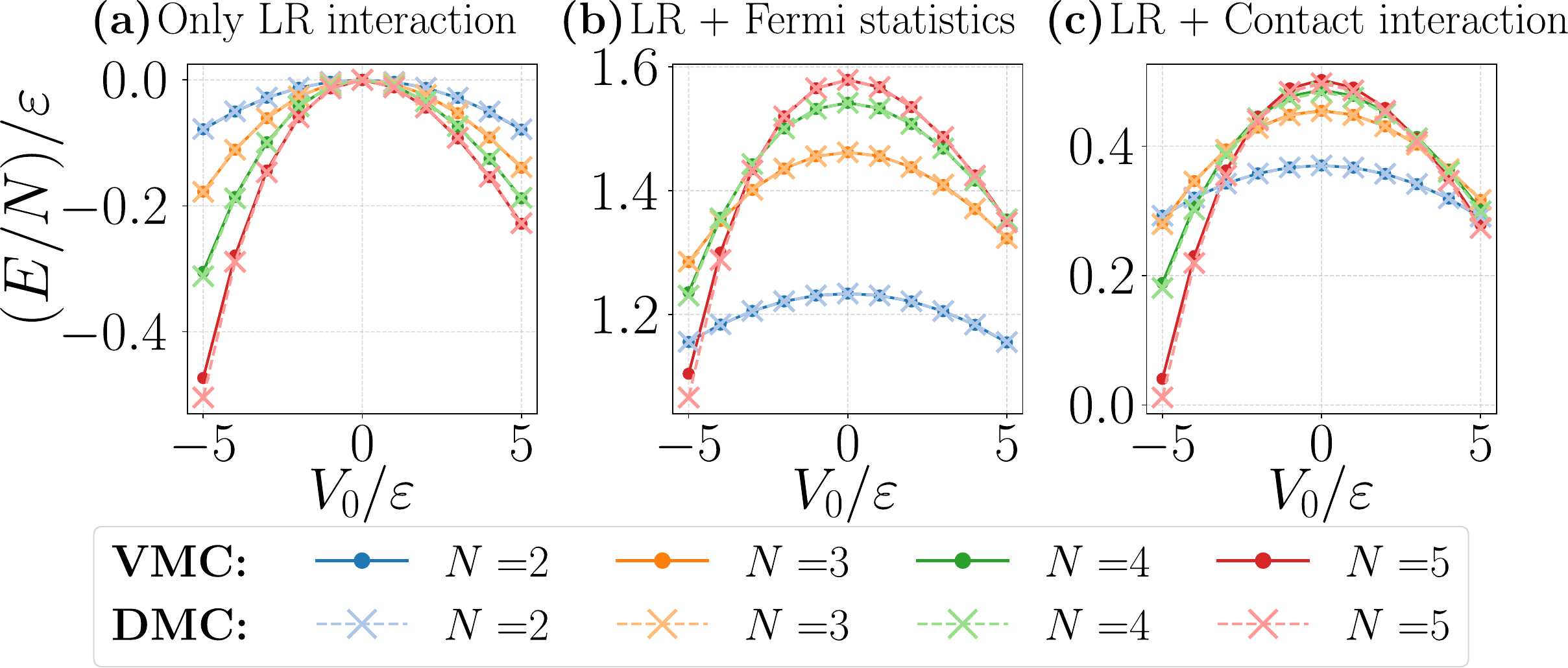}
    \caption{Comparison of the VMC and DMC energies for different statistics and interactions. 
    Overall, there is a close agreement between variational results (circles) and exact ones (crosses). 
    Variational energy is slightly larger according to the variational principle, while close agreement justifies the use of the product form of the trial wave function, summarized in Table~\ref{tab:summary_trial_wf}. 
    Panel (a): Long-range interaction only. Panel (b): Long-range interaction with fermionic statistics. Panel (c): Combined contact and long-range interactions. 
}
    \label{fig:VMC_vs_DMC}
\end{figure} 

Figure~\ref{fig:VMC_vs_DMC} shows the characteristic examples of the comparison of the ground state energies obtained using the VMC and DMC methods for different statistics and interactions. 

The product trial wave function (summarized in Table~\ref{tab:summary_trial_wf}) is used for both methods, serving as the variational ansatz in VMC and the guiding function in DMC calculations. As expected, the DMC energies are systematically lower than the VMC ones, in agreement with the variational principle. The small energy difference shows good quality of the trial wave function and justifies the product form of the trial wave function. 

Overall, the agreement between VMC and DMC is very good in all three cases shown in Fig.~\ref{fig:VMC_vs_DMC}, demonstrating that the trial wave function captures the relevant many-body correlations. The agreement is particularly good for the purely long-range interaction [Fig.~\ref{fig:VMC_vs_DMC}(a)], while slightly larger deviations are observed in the fermionic case  [Fig.~\ref{fig:VMC_vs_DMC}(b)] and in the presence of contact interactions [Fig.~\ref{fig:VMC_vs_DMC}(c)], where they reflect, respectively, the increased importance of short-range correlations and the fixed-node approximation. These results validate the quality of the trial wave function proposed in this work.

\section{Determination of the self-organization onset}
\label{sec:criterion}

\begin{figure}[t]
\centering
\includegraphics[width=\linewidth]{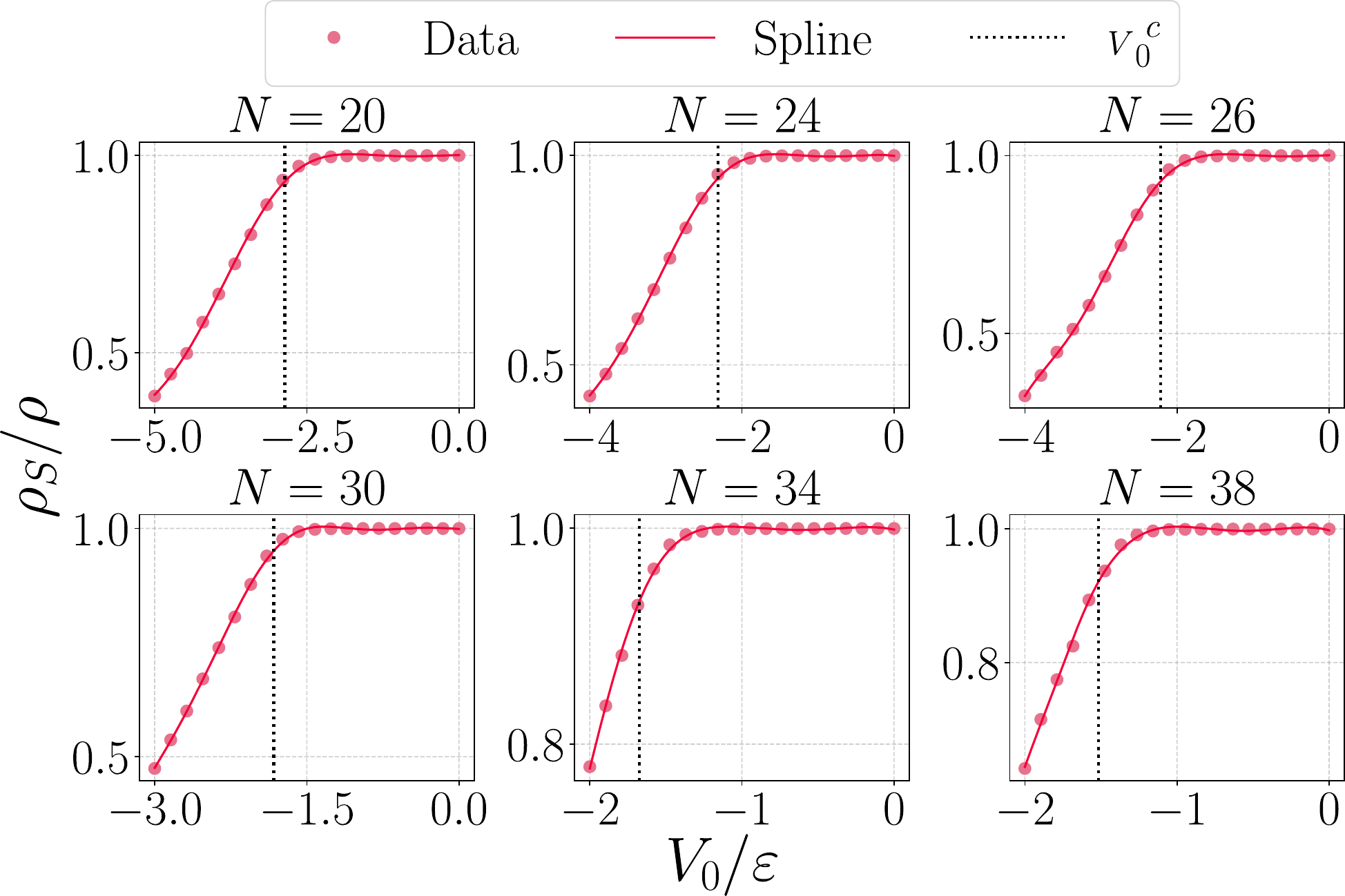}
\caption{
Determination of the critical interaction amplitude $V_0^c(N)$ from the superfluid fraction.
Characteristic examples of the dependence of the superfluid fraction $\rho_S/\rho$ on the amplitude of the long-range interaction $V_0$ 
for the fermionic system at density $\rho = 2\lambda^{-1}$. 
Each panel corresponds to a different particle number $N$. Symbols show the values of the superfluid fraction $\rho_S/\rho$ obtained from the Leggett bound, Eq.~\eqref{eq:legget_bound}, and the solid lines the cubic spline interpolation used to approximate the data and evaluate its derivatives. The vertical dashed lines indicate the critical interaction amplitude $V_0^c(N)$, defined as the point at which the third derivative of the interpolated curve vanishes, $(\rho_S/\rho)'''=0$, within the region where the curve departs from the plateau.
}
\label{fig:rhos_rho_criterion}
\end{figure}

The critical interaction amplitude $V_0^c(N)$ used to construct the phase diagram reported in Fig.~\ref{fig:phase_diagram} is extracted from the superfluid fraction. For each particle number $N$ and each type of short-range repulsion, we compute $\rho_S/\rho$ as a function of $V_0$ from the Leggett bound, Eq.~\eqref{eq:legget_bound}. 
Typically, the system is fully superfluid at weak attraction, while the superfluid response gradually vanishes as the system localizes in the minima of the cavity potential.

Figure~\ref{fig:rhos_rho_criterion} shows representative examples of how $V_0^c(N)$ is obtained for different particle numbers. Starting from the computed values of $\rho_S/\rho$ at discrete values of $V_0$, we interpolate the data with a smoothing cubic spline and evaluate its second derivative, $(\rho_S/\rho)''$. We restrict this analysis to the region where $\rho_S/\rho$ departs from the plateau and decreases towards the self-organized state, excluding the plateau itself. Within this restricted range, $(\rho_S/\rho)''$ has a single, well-defined minimum, and we identify $V_0^c(N)$ as the point at which this minimum occurs. This criterion is applied uniformly across all particle numbers and both types of short-range repulsion. The phase diagram of Fig.~\ref{fig:phase_diagram} is constructed from the values obtained in this way.

\section{Thermodynamic Limit extrapolation}
\label{sec:extrapolation}
 
\begin{figure}[t]
    \centering
    \includegraphics[width=\linewidth]{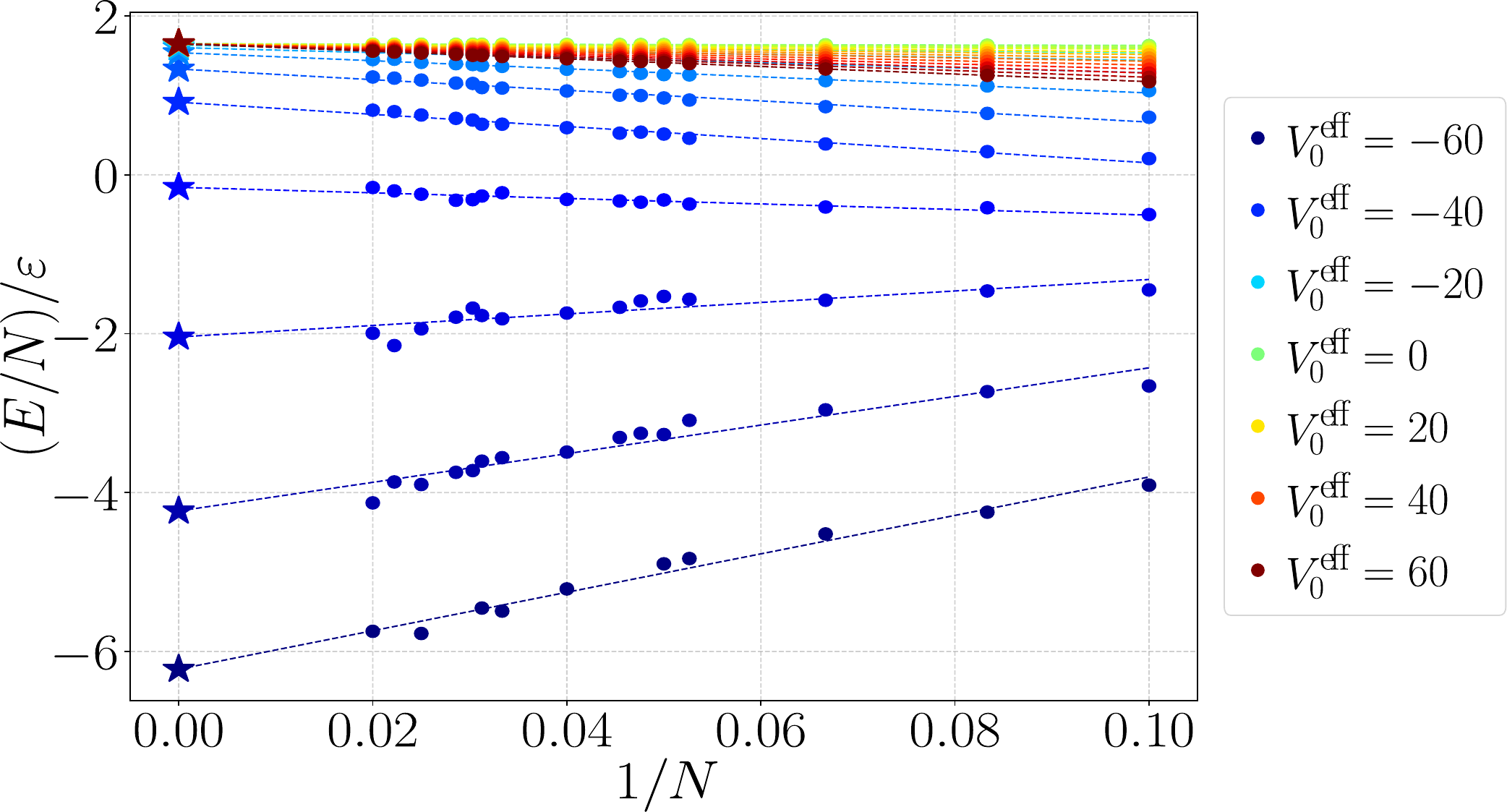}
    \caption{Examples of the linear inverse-number-of-particles $1/N$ extrapolation of the energy per particle $E/N$ to the thermodynamic limit $N\to\infty$ for the fermionic system with long-range interactions for the scaled amplitude of the long-range interactions, $V_0^{\text{eff}} \equiv NV_0 \in\{-60, -55, -50, \dots, 55, 60\}$. Symbols show the Monte Carlo values of the energy per particle $E/N$ as a function of the inverse number of particles $1/N$ at constant density $\rho = 1\lambda^{-1}$, for the indicated values of $V_0^{\text{eff}}$. Dashed lines represent the linear fits, and stars show the obtained thermodynamic ($1/N = 0$) values of the energy, also reported in Fig.~\ref{fig:kac_thermo_all}.}
    \label{fig:1N_EN_extrapolation}
\end{figure}

Figure~\ref{fig:1N_EN_extrapolation} shows examples of the Kac extrapolation procedure for the energy as described in Sec.~\ref{sec:kac-energy}, for the fermionic system with long-range interactions.

For a sufficiently large number $N$ of particles, a linear behavior in $1/N$ is observed for different values of the scaled amplitude of the long-range interactions, $V_0^{\text{eff}} \equiv NV_0$, both for the repulsive and attractive sides. 
The same procedure is applied to bosons with long-range interactions only, and bosons with combined short- and long-range interactions. 

As well, we use a linear in $1/N$ extrapolation procedure to extrapolate the thermodynamic density profiles reported in Sec.~\ref{sec:density_kac}.

\bibliography{biblio}
\end{document}